\documentclass[floats,preprint,prb,aps,draft]{revtex4}
\usepackage{amssymb}
\usepackage{fleqn}
\begin{document}
\preprint{\emph{Submitted to J. Chem. Phys.}}
\title{Semiclassical Mode-Coupling Factorizations of Coherent Nonlinear Optical Response}
\author{Thomas la Cour Jansen and Shaul Mukamel}
\affiliation{Department of Chemistry, 516 Rowland Hall, University of California,\\ Irvine, California 92697-2025, United States}


\date{\today}
\begin{abstract}

The identification of relevant collective coordinates is crucial
for the interpretation of coherent nonlinear spectroscopies of
complex molecules and liquids. Using an $\hbar$ expansion
of Liouville space generating functions, we show how to factorize
multitime nonlinear response functions into products of
lower-order correlation functions of collective coordinates, and derive
closed expressions for linear, second and third order
response functions. In addition to providing systematic quantum corrections,
$\hbar$ offers a convenient bookkeeping
device even for the purely classical response, since including quantum fluctuations allows to circumvent the expensive computation of stability matrices which is a major
bottleneck in Molecular Dynamics (MD) simulations. The existing classical simulation
strategies, including Mode-Coupling in $\mathbf{k}$ space and in real-space,
Langevin equations, and Instantaneous Normal Modes are
compared from a unified viewpoint.

\end{abstract}
\maketitle
\section{Introduction}

In 1993 Tanimura and Mukamel had proposed the fifth order
Raman response~\cite{Tanimura.1993.JCP.99.9496} as a multidimensional spectroscopic
technique especially suitable for investigating the structure and
dynamics of molecular liquids by revealing
detailed information unavailable from linear spectroscopies. That
article had triggered an intense experimental investigations mainly on
liquid $CS_2$.
~\cite{Steffen.1996.PRL.76.1224,Steffen.1997.JCP.106.3854,Steffen.1998.CPL.290.229,Tokmakoff.1998.CP.233.231,Tokmakoff.1997.JCP.106.2569,Tominaga.1996.JCP.104.4419,Blank.1999.JCP.111.3105,Golonzka.2000.JCP.113.9893,Astinov.2000.OL.25.853,Astinov.2000.CPL.327.334,Kubarych.2002.JCP.116.2016,Kubarych.2002.Ultrafast,Kubarych.2003.CPL.369.635,Kaufman.2001.JCP.114.2312,Kaufman.2002.PRL.88.207402}
Earlier experimental investigations were haunted by competing,
sequential, low order, (cascading) processes. Separating the direct and sequential contributions had
drawn considerable attention.~\cite{Ulness.1998.JCP.108.3897,Blank.1999.JCP.111.3105,Kirkwood.1998.CPL.293.417,Kaufman.2001.JCP.114.2312}

Several theoretical methods have been employed to predict the
fifth order response from molecular dynamics simulations of
liquids. Two methods obtain the response directly without further
approximations, other than that the
response is classical. The first, based on calculating time
correlation functions, relies on propagation of the full stability
matrix (Eq. (\ref{eq:stability.matrix})).~\cite{Mukamel.1996.PRE.53.R1,Saito.1998.JCP.108.240,Khidekel.1996.B01}
Since the stability matrix depends on the number of phase space
coordinates squared, it is very time consuming and was only
implemented for very small systems~\cite{Ma.2000.PRL.85.1004,Ma.2002.JCP.116.4962,Saito.2002.PRL.88.207401,Saito.submitted}.
The other, Finite Field, method is based on propagating only one column of the
stability matrix, giving rise to a particular response function of
interest, significantly reducing the computational effort. This
method is a direct simulation of the
experiment, where forces originating from interactions between the
electric fields and the molecules are incorporated in the
simulation on the fly.
\cite{Jansen.2000.JCP.113.307,Jansen.2001.JCP.114.10910,Jansen.2002.B01}
One drawback of computing the actual non
equilibrium response rather than response functions is that the entire the full
simulation needs to be repeated for each choice of time intervals
and pulse configurations. Both of these methods are therefore
computationally very demanding. Developing alternative
numerically more affordable approaches, which could provide physical insight will
therefore be highly desirable.

Some discrepancies currently exist between various real-space simulations
performed under slightly different simulation conditions on liquid
$CS_2$.~\cite{Saito.2002.PRL.88.207401,Jansen.2003.PRB.67.134206}
These differences are most pronounced along the second time axis,
($\tau_{32}$), where both a
ridge~\cite{Jansen.2003.PRB.67.134206,Jansen.2003.BKCS.submitted}
and nodes~\cite{Saito.2002.PRL.88.207401,Saito.submitted} have been reported. A
fundamental understanding of the underlying physical processes
should help resolve the questions about the origin of the nodes and the ridges.

The first approximate scheme employed to analyze the fifth order
Raman response was based on the Instantaneous Normal Modes (INM)
~\cite{Stratt.1995.ACR.28.201,Saito.1998.JCP.108.240,Murry.1998.JCP.109.2814,Murry.1998.JCP.109.7913,Ma.2000.PRL.85.1004,Ma.2002.JCP.116.4972,Keyes.1997.JCP.106.46,Keyes.2000.JCP.112.287,Ji.2000.JCP.112.4186}.
This method uses snapshots of the liquid ``normal modes'' assuming
that they are harmonic and do not change over the timescale of the
experiment. However, in general the normal modes do change on the
timescale of their own periods~\cite{JCP.102-2365} and recent
studies have shown that the INM gives rather poor results for the
fifth order response of liquid Xenon
.~\cite{Ma.2002.JCP.116.4972,Ma.2002.JCP.116.4962}

A more attractive procedure is to identify some relevant
\textit{collective coordinates} and adopt a reduced description for the
response. Unlike microscopic INM where a harmonic model for molecular liquids
may not be
justified, collective coordinates can have Gaussian statistics by
virtue of the central limit theorem.  A simple and tractable
physical picture for the origin of the response is then, in
principle, possible. The multimode Brownian oscillator model has
been successfully employed in the analysis of solvation dynamics in electronic
spectroscopy,~\cite{Chernyak.1998.JCP.108.5812,Mukamel.1995.B01,Mukamel.1999.JCP.110.1711,Khidekel.1996.B01}
where the response may be
expressed using  a few (overdamped or underdamped) collective
coordinates. This model has been used to simulate the fifth order
response of liquid water~\cite{JPC.98-12466,JPC.100-10380}, but
identifying the microscopic origin of these modes still remains an
open challenge~\cite{ACP.84-435}. Nonlinear hydrodynamics and mode
coupling theories successfully use collective variables in
momentum ($\mathbf{k}$) space to describe slow, long-wavelength,
variables and their fluctuations. Mode-Coupling (MC) theory
\cite{Schofield.1992.PA.181.89,Bouchaud.1996.PA.226.243,PA.99-403,Gotze.1989.B01,RPP.55-241,PA.107-25,PRL.50-590}
has been applied to relate the fifth order Raman response
\cite{Denny.2001.PRE.63.065101,Denny.2002.JCP.116.1987,Denny.2002.JCP.116.1979,Cao.2002.JCP.116.3739,Cao.2002.JCP.116.3760}
to fluctuations of density modes is
$\mathbf{k}$ space. Another related approach is based on the Generalized
Langevin Equations (GLE)~\cite{Kim.2002.PRE.65.061102,Zwanzig.2001.B01}.
Classical mode-coupling theory contains some ambiguities regarding
the proper factorization of high order response functions, and
Schofield~\cite{Zon.2002.PRE.65.011106} and
Keyes~\cite{Kim.2002.PRE.65.061102} and collaborators have discussed possible
simulation strategies based on Langevin equations.

In this paper we apply a unifying picture of quantum field and mode-coupling Green function theories
developed recently~\cite{Mukamel.1999.JCP.110.1711,Mukamel.submitted,PR.263-213} to derive expressions for the first-, second- and third order response functions.
The technique provides an unambiguous and unique factorization
scheme of multitime correlation functions and allows the perturbative
incorporation of anharmonicities as well as quantum corrections
through an $\hbar$ expansion. Applications are made to the fifth
order Raman response and compared with other approximate
methods. In Section II we present the superoperator formalism. In
Section III we describe how third and fifth order Raman response
can be obtained from the general first- and second order response
functions. We discuss the connections, similarities and
differences of the present formulation with other approaches.
Conclusions are given in Section IV.

\section{Liouville space formulation of response functions}

The present approach is based on superoperators ($A_+$ and $A_-$)
corresponding to an ordinary Hilbert space operator $A$,
defined by their action on some Hilbert space operator
$\Omega$.~\cite{Chernyak.1998.JCP.108.5812}

\begin{eqnarray}\label{eq:superoperators}
A_+\Omega & \equiv & \frac{1}{2}(A\Omega+\Omega A), \nonumber\\
A_-\Omega & \equiv & A\Omega-\Omega A.
\end{eqnarray}
Using this notation, A compact expression for the $n$ dimensional $(nD)$ quantum
mechanical response functions can then be written in terms of the
dipole superoperators ($\mu_+$ and
$\mu_-$).~\cite{Mukamel.submitted,Chernyak.1996.JCP.105.4565,PR.263-213}
\begin{equation}
R^{(nD)}=\left(\frac{i}{\hbar}\right)^n\left\langle\mu_+
(\tau_{n+1})\mu_-(\tau_n)\cdots\mu_-(\tau_1)\right\rangle.
\end{equation}
The superscript of $R^{(nD)}$ indicates that it depends on the $n$
time intervals between successive times $\tau_{1}\cdots
\tau_{n+1}$ and thus constitutes an $n$ dimensional technique. The
average $\langle A \rangle \equiv \textrm{Tr}[A \rho_{eq}]$ denotes the trace with respect to the equilibrium density
matrix of the system $\rho_{eq}$.

The Hamiltonian will be partitioned into a Harmonic, quadratic, part ($H_{0}$) and
an anharmonic ($V$) part
\begin{eqnarray}
H=H_{0}+V,
\end{eqnarray}
and the response function of a weakly anharmonic system will be
expanded perturbatively in $V$.~\cite{Mukamel.submitted}
\begin{eqnarray}\label{responsefunction}
R^{(nD)}&=&\sum_{m=0}^{\infty}\left(\frac{i}{\hbar}\right)^{m+n}(-1)^m
\int_{-\infty}^{\tau_{n+1}}d\tau'_1\cdots\int_{-\infty}^{t_{n+1}}d\tau'_m\nonumber\\
&\times&\left\langle T\tilde{\mu}_+(\tau_{n+1})\tilde{\mu}_-(\tau_n)\cdots\tilde{\mu}_-(\tau_1)
\tilde{V}_-(\tau'_m)\cdots \tilde{V}_-(\tau'_1)\right\rangle_0.
\end{eqnarray}
Here $\tilde{V}_{\nu}$ and $\tilde{\mu}_{\nu}$ are the
superoperators associated with the anharmonic part of the
potential and with the interaction dipole, respectively in the interaction
picture with respect to $H_{0}$, i.e.,
\begin{eqnarray}
\tilde{A}_{\nu}(\tau) & \equiv &
\exp\left(\frac{i}{\hbar}L_0\tau\right)A_{\nu}\exp\left(-\frac{i}{\hbar}L_0\tau\right).
\end{eqnarray}
Here
\begin{eqnarray}
L_0\Omega  & = & [ H_0 , \Omega ] \equiv (H_0)_- \Omega,
\end{eqnarray}
is the Liouville operator corresponding to $H_0$ . The average
$\langle A \rangle_0 \equiv \textrm{Tr} [ A \rho_0 ]$ is defined
as the trace with respect to the equilibrium density operator for
the harmonic system $\rho_0$. $T$ is the time ordering operator in
Liouville space which arranges all superoperators so that their time
arguments decrease from left to right~\cite{Mukamel.submitted}.

We shall represent $H_0$ in terms a few primary (collective)
coordinates $Q_j$ 
\cite{Mukamel.1999.JCP.110.1711,Chernyak.1996.JCP.105.4565,PA.121-587}
described by the Hamiltonian
\begin{equation}\label{2.1t}
  H_m = \sum_j \left (\frac {P_j^2} {2M_j} + \frac {M_j \Omega_j^2Q_j^2}
   {2} \right )  + V({\bf Q}),
\end{equation}
where $P_j(Q_j)$ is the momentum (coordinate) operator of the j'th
primary mode, $\Omega_j$ and $M_j$ are its frequency and reduced
mass respectively.
The anharmonic potential $V$, is
\begin{equation}
V({\bf Q})=\sum_{N=3}^{\infty}\frac{1}{N!}V^{(N)}_{j_1\cdots j_N}Q_{j_1}\cdots Q_{j_N}.
\end{equation}
We assume that the dipole operator $\mu$ only depends on the primary
coordinates and expand it as
\begin{equation}\label{eq:alpha.taylor}
\mu=\sum_{N=1}^{\infty}\frac{1}{N!}\mu^{(N)}_{j_1\cdots j_N}Q_{j_1}\cdots Q_{j_N}.
\end{equation}

The primary modes further interact with a large number
of low-frequency harmonic (bath) coordinates which induce relaxation and
dephasing. These bath degrees of freedom {\bf q} and
their linear coupling to the primary modes are described by the
Hamiltonian $H_B$.
\begin{equation}\label{2.7t}
  H_B = \sum_{j\alpha} \left [ \frac {p_{j\alpha}^2} {2m_{j\alpha}}
      + \frac {m_{j\alpha} \omega_{j\alpha}^2} {2}
      \left(q_{j\alpha} - \frac {c_{j\alpha}} {m_{j\alpha} \omega_{j\alpha}^2}
       Q_j\right)^2 \right ],
\end{equation}
and $p_{j\alpha}(q_{j\alpha})$ are momentum (coordinate) operators
of bath oscillators. $c_{j\alpha}$ are the coupling constants between the
primary and bath coordinates.
The total harmonic Hamiltonian is given by
$^{\cite{Mukamel.1995.B01}}$
\begin{equation}\label{2.6t}
H_{0} = H_m ({\bf Q}) +  H_B ({\bf Q}, {\bf q}).
\end{equation}

Applying the algebraic rules for the superoperators given in Appendix A
\cite{Mukamel.submitted} the Taylor expansions of $\mu_{+}(\tau), \mu_{-}(\tau)$ and
$V_{-}(\tau)$ can be expressed in terms of the elementary
superoperators $Q_{j+}$ and $Q_{j-}$.
Using these expansions,  we can convert the
time-ordered product of superoperators into a time ordered product
of primary superoperators. This transforms the computation of
response functions (Eq.~(\ref{responsefunction})) to the
evaluation of products of the form
\begin{equation}
\label{63a} W \{j_{m}\nu_{m}\tau_{m}\} \equiv \langle
Q_{j_1\nu_1}(\tau_1)\ldots Q_{j_N \nu_N}(\tau_N)\rangle{}_0,
\end{equation}
where $\nu_{1,\ldots,}$ $\nu_N=\pm$, and $j_{m}$ runs over the
collective coordinates. Note that the number $N$ of operators in
the product needed to compute $R^{(nD)}$ is generally greater than
$n, N \geq n$.  The reasons are (i) $\mu_{\nu}$ may be nonlinear
in the elementary operators. (ii) The expansion in $V_{-}$ adds
more operators to the product. Generally some of the $\tau_{j}$ in
Eq.~(\ref{63a}) will be the same since $R^{(nD)}$ only depends on
$n+m+1$ times, which is smaller than or equal to $N$.

We next introduce the \emph{superoperator generating
functional}\cite{Mukamel.submitted}
\begin{eqnarray}
S (\{J (t)\})\equiv \left\langle T\exp\left[\sum_{j\nu}\int J_{j
\nu}(\tau) Q _{j \nu} (\tau) d\tau\right]\right\rangle_{0}.
\end{eqnarray}

Since the Hamiltonian $H_{0}$ is quadratic, the generating
functional may be computed exactly using the second order cumulant
expansion. This gives
\begin{eqnarray}
\label{Wick} && S (\{J (t)\})= \exp \left\{
\sum_{j,k}\int_{-\infty}^{\infty}d \tau_{2}
\int_{-\infty}^{\tau_{2}}d \tau_{1}  \right. \\\nonumber &&
[-i{\hbar} J _{j+}(\tau_{2}) J _{k-}(\tau_{1}) \left.
G_{jk}^{+-}(\tau_{21}) + J _{j+}(\tau_{2}) J
_{k+}(\tau_{1}) G_{jk}^{++}(\tau_{21})] \right.\Bigg \},
\end{eqnarray}
We have introduced the notation
$\tau_{ij}\equiv\tau_i-\tau_j$ and the two basic Liouville space Green
functions.
\begin{eqnarray}\label{green}
G_{ij}^{+-}(\tau_{21}) & \equiv & \frac{i}{\hbar} \left\langle T
Q^i_+(\tau_2) Q^j_-(\tau_1) \right\rangle_0,
\end{eqnarray}
\begin{eqnarray}\label{green2}
G_{ij}^{++}(\tau_{21}) & \equiv & \left\langle T Q^i_+(\tau_2)
Q^j_+(\tau_1) \right\rangle_0.
\end{eqnarray}
Using the Eq.~(\ref{eq:superoperators}),
Eqs.(\ref{green}) and (\ref{green2}) can be recast as
combinations of ordinary (Hilbert space) correlation functions
\begin{eqnarray}
G_{ij}^{+-}(\tau_{21}) & = & \theta(\tau_{21})\frac{i}{\hbar} (\left\langle
Q^i(\tau_2) Q^j(\tau_1) \right\rangle_0-\left\langle
Q^j(\tau_1) Q^i(\tau_2) \right\rangle_0), \\
G_{ij}^{++}(\tau_{21}) & = & \frac{1}{2} (\left\langle
Q^i(\tau_2) Q^j(\tau_1) \right\rangle_0+\left\langle Q^j(\tau_1)
Q^i(\tau_2) \right\rangle_0).
\end{eqnarray}
$G^{++}$ and $G^{+-}$ may also be expressed in terms of the
spectral densities $C_{ij}(\omega)$
\begin{eqnarray}\label{eq:spectraldensity}
G^{+-}_{ij}(\tau) & = & 2 \theta(\tau)\int_{-\infty}^ {\infty}\frac{d
\omega}{2 \pi} C_{ij}(\omega)\sin(\omega \tau), \\
G_{ij}^{++}(\tau)
& = & \hbar\int_{-\infty}^{\infty}\frac{d \omega}{2 \pi}
C_{ij}(\omega)\cos(\omega \tau)\coth\left(\frac{\hbar \omega}{2 k_b
T}\right).
\end{eqnarray}
Eq.~(\ref{eq:spectraldensity}) is the definition of the spectral
density, where $\theta(\tau)$ is the Heaviside step function (equal
to zero for $\tau < 0$ and equal to one for $\tau\geq 0$) . In the
classical, high temperature, limit $\coth(\hbar\omega/2k_BT)\approx 2k_BT/\hbar\omega$, and both $G^{+-}$ and $G^{++}$
become independent of $\hbar$. In that case the two are
related by the classical fluctuation-dissipation relation
\begin{equation}
G^{+-}_{ij}(\tau)=-\frac{1}{k_BT}\theta(\tau)\frac{d}{d \tau}
G^{++}_{ij}(\tau).
\end{equation}
Semiclassical approximations to the response may be developed by
expanding $G^{++}$ in powers of $\hbar$.

Time-ordered correlation functions of superoperators may be
obtained from the generating functional by functional
derivatives~\cite{Mukamel.submitted}
\begin{eqnarray}
\label{68}
 W \{j_{m}\nu_{m}\tau_{m}\}=\frac{\partial}{\partial
J_{j_{1}\nu_{1}}(\tau_{1})}\ldots \frac{\partial}{\partial
J_{j_{N}\nu_{ N}}(\tau _{N})} S \{J(\tau)\} \bigg|_{J=0}.
\end{eqnarray}

In order to compute the response function (which gives the
response to very short pulses) to a given order in the field, the
generating functional can be simplified since only a limited
number of times will contribute, and the primary operators
connected with the last time will have to be $+$ operators since
$\langle A_{-} (\tau_{1})B_{\nu}(\tau_{2})\rangle$ vanishes for
$\tau_1 \geqslant \tau_2$. ~\cite{Mukamel.submitted}
The generating functional for two time
quantities $\langle A_{\nu} (\tau_{1})B_{\nu'}(\tau_{2})\rangle$
thus reads
\begin{eqnarray}
 S^{(1)} (\{J(\tau) \}) &=& \exp \bigg\{
\sum_{j,k}-i\hbar J _{j+}(\tau_{2}) J _{k-}(\tau_{1})
G_{jk}^{+-}(\tau_{21})+J _{j+}(\tau_{2}) J
_{k+}(\tau_{1}) G_{jk}^{++}(\tau_{21}) \nonumber\\
&+& J _{j+}(\tau_{2}) J _{k+}(\tau_{2}) G_{jk}^{++}(0) +J
_{j+}(\tau_{1}) J _{k+}(\tau_{1})
G_{jk}^{++}(0)\bigg\}.
\end{eqnarray}
 Here we assumed Gaussian statistics of $Q_{j}$ so that the exact generating
functional is given by the second order cumulant expansion. The generating functional provides a compact form for Wick's theorem.
 Generating functionals for multitime quantities may be written in a similar way.

Using the general expression for the $n$'th order response function given in Appendix A, we can expand $R^{(nD)}$ to any desired
order in the primary operators. Since the $J$'s always come in
pairs in the generating functional, the derivatives will vanish
for all terms with an odd number of elementary operators, once the
$J=0$ limit is taken.

Note the delicate interplay of the $\hbar$ factors in
Eq.~(\ref{eq:monster}). Keeping $\hbar$ alive even in the
classical limit, is what allows us to avoid the computation of
stability matrices.  $\hbar$ retains information about quantum fluctuations which are differences between ``left'' and ``right''
trajectories~\cite{Mukamel.1996.PRE.53.R1,Khidekel.1996.B01,Chernyak.1998.JCP.108.5812}:
The stability matrices are the corresponding derivatives as
$\hbar$ tends to zero. In the classical evaluation of
Eq.~(\ref{eq:monster}) we keep terms in the generating function to
order ($\hbar^{n+m})$. $\hbar$ then cancels out by the prefactor in Eq. (\ref{responsefunction})
and the result is independent of $\hbar$, as it should be. \cite{Chernyak.1998.JCP.108.5812}  Higher
order terms in $\hbar$ provide quantum corrections to the
response.  We further note that the response function, which
is a particular combination of correlation functions, has a
well-defined classical limit. Individual correlation functions in
Hilbert space generally do not have a clear physical meaning and
consequently their $\hbar\rightarrow 0$ limit is ill defined.
$\hbar$ therefore cancels only once these combinations are
evaluated to yield an observable.

As an example, the $1D$ response function expanded to fourth order in the
elementary operators is given by
\begin{eqnarray}\label{eq:firstorder}
R^{(1D)}&=&\sum_{ij}\mu^{(1)}_i\mu^{(1)}_j G_{ij}^{+-}(\tau_{21})\nonumber\\
&+&\sum_{ijkl}\mu^{(2)}_{ij}\mu^{(2)}_{kl} G^{+-}_{ik}(\tau_{21})G^{++}_{jl}(\tau_{21})+\sum_{ijkl}\mu^{(3)}_{ijk}\mu^{(1)}_{l}
G^{+-}_{il}(\tau_{21})G^{++}_{jk}(0).
\end{eqnarray}
These two lowest-orders in the expansion are independent of the
anharmonicity and the $1D$ response can be expected to be
dominated by the harmonic part of the potential.
Closed expressions for the 2D and 3D response functions expanded to
sixth order in the primary coordinates are given in Appendix B.

\section{Comparison of simulation strategies for 2D fifth order Raman response}
In off-resonant Raman spectroscopy, the field-matter
interaction comes through the dipole moment induced by an electric
field instead of the permanent dipole moment. The Raman response
can therefore be obtained by simply substituting the dipole operators in
the expressions in Eq. (\ref{eq:firstorder}) and Appendix B with the operator for the
induced dipole,
which in turn is given by the product of the polarizability and the
inducing field. $\mu$ is therefore simply replaced by $\alpha\cdot
E$. This substitution leads to a higher order dependence on the
electric field: The $nD$ response is n'th order in the field for dipole (e.g.
infrared) response but $(2n+1)$'th order for Raman. The Raman response that is third
order in the electric fields is therefore described by the 1D response function.
The fifth order Raman response is given by the $2D$ response function and so forth.

The various approaches used for the simulation of fifth-order Raman signals differ not only by the simulation technique, but also by the
model used, which complicates their direct comparison. The expressions
derived in this paper allow a critical comparison of the various simulations. This will be done next.

\underline{Classical MD simulations in real-space:}

The real-space simulations by Saito \textit{et
al.}~\cite{Saito.2002.PRL.88.207401}, Ma \textit{et
al.}~\cite{Ma.2000.PRL.85.1004,Ma.2002.JCP.116.4962}  and Jansen
\textit{et
al.}~\cite{Jansen.2000.JCP.113.307,Jansen.2001.JCP.114.10910} use
the nuclear coordinates as basis and the response is calculated
without invoking Wick's theorem. The calculated classical response
functions therefore formally include all orders of the expansion
derived here. The price is the need to compute stability matrices.
The nuclear coordinates might not be the best choice of basis for
the expansion derived in the previous section and analyzing the response
in this basis might be inconvenient.
The real-space simulations give the full response
including all anharmonicities and nonlinearities of the
polarizability. Analyzing the response calculated with these
methods is a daunting task, since all terms which depend on
different modes and nonlinearities are added. This makes it difficult to
establish the connection between the spectral features and
the underlying dynamics.

The stability matrix ($M(\tau_2,\tau_1)$) is a $N$ by $N$ matrix, where $N$ is the number of phase space coordinates. \cite{Mukamel.1996.PRE.53.R1,Saito.1998.JCP.108.240,Dellago.2003.PRE.67.035205,Ma.2002.JCP.116.4962} Each matrix element is given by the derivative of a phase space coordinate $x_k$ at time $\tau_1$ with respect to a phase space coordinate $x_j$ at another time, $\tau_2$
\begin{equation}\label{eq:stability.matrix}
M_{jk}(\tau_2,\tau_1)=\left\{\frac{\partial x_k(\tau_1)}{\partial x_j(\tau_2)}\right\}.
\end{equation}
Schemes for obtaining the stability matrix using the Hessian matrix have been described in the literature.~\cite{Mukamel.1996.PRE.53.R1,Saito.1998.JCP.108.240,Ma.2002.JCP.116.4962,Dellago.2003.PRE.67.035205,Brack.1997.B01.C,Williams.2000.JCP.113.10651,Saito.submitted}

Equilibrium simulations
~\cite{Saito.2002.PRL.88.207401,Ma.2000.PRL.85.1004,Ma.2002.JCP.116.4962,Mukamel.1996.PRE.53.R1,Saito.1998.JCP.108.240,Saito.submitted}
are based on propagating the full stability matrix
\cite{Mukamel.1996.PRE.53.R1} in order to evaluate the Poisson
bracket arising in the time correlation function expression for
the fifth-order response. The major bottleneck in the equilibrium method is the propagation of the $N\times N$ stability matrix.~\cite{Mukamel.1996.PRE.53.R1,Saito.1998.JCP.108.240} In contrast, in the non equilibrium, Finite Field, approach
~\cite{Jansen.2000.JCP.113.307,Jansen.2001.JCP.114.10910,Dellago.2003.PRE.67.035205},
propagation of the full stability matrix is avoided and only a few vectors
of dimension $N$ corresponding to the response function are propagated. In this
approach the evaluation of the first order derivatives of the polarizability, needed in order to calculate the forces exerted by the electric fields, is the most
time consuming part of the simulation.~\cite{Jansen.2001.JCP.114.10910}
 
\underline{Instantaneous Normal Modes:}

For a collection of oscillators with
frequencies $\omega_{i}$ we have \cite{Tanimura.1993.JCP.99.9496}
\begin{eqnarray}\label{eq:INM}
G^{+-}_{ij}(\tau) & = & \theta(\tau)\frac{1}{M_{i}\omega_{i}}\sin(\omega_{i} \tau)\delta_{ij} \nonumber \\
G^{++}_{ij}(\tau)
& = & \hbar\frac{1}{2 M_i\omega_{i}}\cos(\omega_{i} \tau)\coth\left(\frac{\hbar \omega_{i}}{2 k_b T}\right)\delta_{ij}.
\end{eqnarray}
INM simulations combine these expressions with MD simulations
of real liquids. The system is assumed to evolve
independently in the different normal modes giving rise to
the Kronecker deltas in Eq. (\ref{eq:INM}). 
INM simulations have most often been applied
in the classical, high temperature limit, where only the
first term in the expansion of $\coth$ is retained
$(\coth(x)\approx 1/x)$. 
Eqs.~(\ref{eq:fifthorder})-(\ref{eq:3111}) reduce to the INM
expressions if the INM coordinates are used as the
primary operators and the anharmonicities are neglected.
In this paper we use a different bookkeeping:
terms have been kept to a
certain order in the primary operators, whereas the INM
expressions traditionally have been truncated by neglecting terms
containing third- and higher order nonlinearities of the
polarizabilty
~\cite{Saito.1998.JCP.108.240,Ji.2000.JCP.112.4186,Ma.2000.PRL.85.1004,Keyes.2000.JCP.112.287},
retaining only Eqs.~(\ref{eq:fourthorder}) and
(\ref{eq:firstsixthorder}). Our formulation suggests that
terms which include second order derivatives should be considered
on an equal footing with all other terms including a total of six
derivatives, since they depend on the same number of fundamental
quantum Green functions $G^{++}$ and $G^{+-}$.

In a recent INM simulation of Ma and Stratt
\cite{Ma.2002.JCP.116.4972} the lowest order anharmonic
contribution corresponding to Eq. (\ref{eq:3111}) was 
found to give a significant contribution to the total response
of liquid Xenon. The fifth-order Raman response computed with this
model did, however, still deviate significantly from that
calculated using time correlation functions.~\cite{Ma.2002.JCP.116.4972,Ma.2002.JCP.116.4962}

Both the diagonalization of the Hessian needed to obtain the normal modes and
the evaluation of the derivatives of the polarizability are time consuming processes.
Which of these is the slowest depend on the model used to describe the polarizability. For the first order dipole induced dipole model the derivatives can be
evaluated very effectively analytically. This simple model is, however, known to fail for systems with large polarizability \cite{Geiger.1987.JCP.87.191}
and in atomic liquids it gives an isotropic polarizability that is coordinate independent.

In liquid phase, the basic assumption in the INM theory that motion can be described in static normal modes is not generally justified. In a liquid the molecules rotate and diffuse around changing the normal modes on the timescale that is studied in femto and picosecond experiments. Instead of using a static set of coordinates one can use a dynamic basis set. Ma and Stratt used a basis of normal modes that was allowed to change with time. \cite{Ma.2002.JCP.116.4972}
They obtained the Green's function describing the time evolution in one mode by applying WKB theory.\cite{Bender.1978.B01,Ma.2002.JCP.116.4972} In our notation, this result reads 
\begin{equation}
G^{+-}_{ij}(\tau)=\theta(\tau)\frac{\sin\left(\int_{0}^{\tau}\omega_i(\tau')d \tau'\right)}{M_i\sqrt{\omega_i(\tau)\omega_i(0)}}\delta_{ij}.
\end{equation} 
The time evolution depends on the time dependent frequency $\omega_i(\tau)$ or each normal mode. In this picture not only the frequencies are time dependent, but the polarizability derivatives and anharmonicities also change as the basis set change. 
Computationally it is expensive to repeatedly diagonalize the Hessian at short intervals to obtain  the dynamic normal mode basis set and the method was not yet tested.

\underline{Mode-coupling in $\mathbf{k}$ space:}

Reichman
~\cite{Denny.2001.PRE.63.065101,Denny.2002.JCP.116.1987,Denny.2002.JCP.116.1979}
and Cao~\cite{Cao.2002.JCP.116.3739,Cao.2002.JCP.116.3760}
adopted Mode-Coupling theory of space nonlinear hydrodynamics in $\mathbf{k}$. Translational invariance then
greatly simplifies the final expressions and the time dependence
is accounted for through the dynamical structure factor $F(\mathbf{k},t)$.
They employed the atomic first order dipole induced dipole model for
the polarizabilit.
The resulting classical expression is~\cite{Denny.2001.PRE.63.065101,Denny.2002.JCP.116.1987}
\begin{eqnarray}
R^{(2D)}(\tau_{32},\tau_{21})&=&\sum_k\left( \frac{V(\mathbf{k})}{S(\mathbf{k})^2} \right)^3
\frac{1}{k_bT}\frac{d F(\mathbf{k},\tau_{32})}{d \tau_{32}}\nonumber\\
&\times&\left[\frac{1}{k_bT}\frac{dF(\mathbf{k},\tau_{32}+\tau_{21})}{d \tau_{21}}F(\mathbf{k},\tau_{21})+\frac{1}{k_bT}\frac{dF(\mathbf{k},\tau_{21})}{d \tau_{21}}F(\mathbf{k},\tau_{31})\right].
\end{eqnarray}
This result neglects the leading (fourth order) term (Eq.~(\ref{eq:fourthorder})) and only retains one of the sixth order terms (Eq.~(\ref{eq:firstsixthorder})).
The time derivative of the dynamical structure factor is proportional to the Fourier transform of $G^{+-}$, while the dynamical structure factor itself is proportional to the Fourier transform of $G^{++}$. The $V(\mathbf{k})/S(\mathbf{k})^2$ factors correspond to polarizability derivatives in $\mathbf{k}$ space.~\cite{Denny.2001.PRE.63.065101,Denny.2002.JCP.116.1987}  

The quantum mechanical response
function contains information about the local response assuming
that the relevant coherence size is much smaller than the
wavelength of the light. This is obviously the case for the
response of many polyatomic molecules and aggregates. However, it
usually holds for molecular liquids as well, where the coherence
size underlying the response is small due to local disorder. It fails
near critical points where the coherence lengths become very
large.~\cite{PRL.50-590,JOSAB.3-1124,PRA.33-1099,ACP.70-165}
Since the response is local, this expansion should
be made in real-space and not in $\mathbf{k}$ space. The
simulation should provide a localized response since all six
dipoles contributing to $R^{(2D)}$ should act within a small region (of order of the first solvation
shell in $CS_{2})$) in order to generate the local response. Of
course, the signal should eventually be calculated in $\mathbf{k}$
space. This can be done purely macroscopically by solving
Maxwell's equations within the local field approximation
as described in chapter 16 of reference {39}. The only microscopic
information that enters the signal is the effective local
response, which can be obtained from simulations performed on the
entire liquid. 

\underline{Mode-Coupling in real-space; Langevin equations:}

Using the Hamiltonian (Eqs. (\ref{2.1t})-(\ref{2.7t})) the Brownian oscillator motion is described by the following generalized Langevin equation. \cite{Mukamel.1995.B01}
\begin{equation}\label{eq:GLE}
M_j\ddot{Q}_j(t)+M_j\Omega_jQ_j(t)+M_j\sum_i\int^{t}_{-\infty}d\tau
[\gamma_{ji}(t-\tau)+i\Sigma_{ji}(t-\tau)]\dot{Q}_i(\tau)=f_j(t)+F_j(t)
\end{equation}
$\gamma_{ij}$ ($\Sigma_{ij}$) is the imaginary (real) parts of a
self energy operator representing relaxation (level shift). $f_j$ is a Gaussian stochastic random force representing the bath degrees of freedom on the coordinate $j$ and $F_j$ is an external driving force.
\begin{eqnarray}\label{ghi1}
\gamma_{ij} (\omega) & = & \frac{\pi}{M_{i}}\sum_{\alpha} 
\frac{c_{j\alpha}c_{i\alpha}}{2 m_{\alpha}\omega_{\alpha}^2} [
\delta(\omega-\omega_{\alpha}) + \delta(\omega+\omega_{\alpha})],
\end{eqnarray}
$\Sigma$ is related to $\gamma$ by the Kramers-Kronig relation.
\begin{eqnarray}
 \Sigma_{ij}(\omega) & = &-\frac{1}{\pi}{\text Re }\int_{-\infty}^{\infty}d\omega'
 \frac{\gamma_{ij}(\omega')}{\omega'-\omega},
\end{eqnarray}
In Appendix C the matrix of spectral densities is derived by solving Eq.(\ref{eq:GLE}).~\cite{PA.121-587,Mukamel.submitted}
\begin{eqnarray}\label{ghi}
{C''}(\omega) & = & \textrm{Im}\left(\frac{1}{{M}
({\Omega}^2 +\omega{\Sigma} (\omega)-{I}{\omega}^2 +  i\omega\mathbf{\gamma}(\omega))}\right).
\end{eqnarray}
${M}$, ${\Omega}$ and ${I}$ are all diagonal matrices with matrix elements are given as follows $M_{ij}=\delta_{ij}M_j$, $\Omega_{ij}=\delta_{ij}\Omega_j$ and $I_{ij}=\delta_{ij}$.
$C''(\omega)$ is the odd part of the spectral density, which is related to the even part $C'(\omega)$ by the fluctuation-dissipation theorem. \cite{Chernyak.1998.JCP.108.5812} The matrix of spectral densities is given by $C(\omega)=[1+\coth(\beta\hbar\omega/2)]C''(\omega)$, where  
\begin{eqnarray}
{C''}(\omega) & = & \frac{1}{\gamma(\Omega^2-\omega^2I+\Sigma\omega)\gamma^{-1}(\Omega^2-\omega^2I+\Sigma\omega)+\gamma^2\omega^2}\gamma\omega M^{-1}
\end{eqnarray}

Ordinary Langevin equations are obtained by first neglecting the
frequency dependence of $\gamma_{ij}$.
Setting
$\gamma_{ij}(\omega)=\gamma_{ij}$ and $\Sigma_{ij}(\omega)=0$.
We then take the overdamped
limit $\gamma_{ij}>>\Omega_{ij}$ of Eq.~(\ref{ghi}) where the matrix of spectral
densities assumes the form~\cite{Chernyak.1996.JCP.105.4565}
\begin{equation}\label{41}
C''(\omega)=
\frac{1}{\omega^2+\Lambda^2}\Lambda\lambda\omega,
\end{equation}
with the $N\times N$ matrices $\Lambda$ and $\Omega$ are $\Lambda=\gamma^{-1}\Omega^2$ and $\lambda=M^{-1}(\Omega^{-1})^2$,
~\cite{Mukamel.1995.B01,Chernyak.1996.JCP.105.4565} $N$ being the number of Brownian oscillator modes.  
The nonlinear response of systems described by Langevin equations can
be obtained by using Eq.~(\ref{41}) for
the matrix of spectral densities.
Upon the substitution of Eq.~(\ref{41}) in Eq.~(\ref{eq:spectraldensity}) we get for the superoperator Greens functions in matrix notation
\begin{eqnarray}
G^{+-}(\tau) & = & 2 \theta(\tau) \exp(-\Lambda\tau) \Lambda\lambda\\ \label{eq:langevinpp}
G^{++}(\tau)
& = & \hbar  \exp(-\Lambda\tau)\Lambda\lambda\coth(\beta\hbar\Lambda)
+\frac{4}{\beta}
\sum_{n=1}^{\infty}\frac{\nu_n\exp(-\nu_n \tau)}{\nu_n^2-\Lambda^2}\Lambda\lambda
\end{eqnarray}
where $\nu_n\equiv{2\pi n}/{\hbar\beta}$ are the Matsubara frequencies. \cite{Weiss.1993.B01}
In the high temperature ($\beta\hbar\Lambda<<1$) limit the second sum in
Eq.~(\ref{eq:langevinpp}) decays rapidly and may be neglected and the $\coth(\beta\hbar\Lambda)$ factor can be approximated by $1/\beta\hbar\Lambda$
in the first term, yielding
\begin{eqnarray}
G^{++}(\tau)
& = & \beta^{-1}\exp(-\Lambda\tau)\Lambda\lambda\Lambda^{-1}
\end{eqnarray}

In the Generalized Langevin Equation approach of Kim and Keyes \cite{Kim.2002.PRE.65.061102} the response function is factorized using the scheme suggested by van Zon and Schofield \cite{Zon.2002.PRE.65.011106} which assumes fast decay of the fluctuating forces. This factorization results in the first order term for the 2D response proportional to $G^{+-}(\tau_{32})G^{+-}(\tau_{21})$
(the second term in Eq.(\ref{eq:fourthorder})), whereas the first term
$G^{+-}(\tau_{32})G^{+-}(\tau_{31})$ is neglected.
The systematic mode-coupling factorization of the present paper is unambiguous and require no further assumptions about the behavior of the correlation functions.

\underline{Adiabatic simulations:}

In systems with large scale variation of structure such as proteins and liquids it can be useful to employ a dynamic basis set instead of a static one. The natural starting point for such a description is to employ the adiabatic basis.~\cite{Shapere.1989.B01,Griffith.1995.B01,Sakurai.1978.B01}
A quantum description is obtained when one uses the eigenstates of the vibrational Hamiltonian as a dynamic basis. Using this basis we can derive an adiabatic theory for the response functions.
When this basis changes slowly the adiabatic approximation can be employed, considerably simplifying the calculation of the time evolution, since the system remains in the same adiabatic state at all times.

There are two independent Liouville space pathways contributing to the 2D response. \cite{Mukamel.1995.B01}
\begin{equation}
R^{(2D)}(\tau_{32},\tau_{21})=R_1(\tau_{32},\tau_{21})+R_2(\tau_{32},\tau_{21})+c.c.
\end{equation}
In the adiabatic approximation these are given by
\begin{eqnarray}\label{eq:third.order.response.functions}
R_1(\tau_{32},\tau_{21})&=&\left(\frac{i}{\hbar}\right)^2\sum_{abc}
\mu_{ca}(\tau_3)I_{ca}(\tau_{32})\mu_{cb}(\tau_2)I_{ba}(\tau_{21})\mu_{ba}(\tau_1)P(a),
\nonumber\\
R_2(\tau_{32},\tau_{21})&=&\left(\frac{i}{\hbar}\right)^2\sum_{abc}
\mu_{cb}(\tau_3)I_{bc}(\tau_{32})\mu_{ca}(\tau_2)I_{ba}(\tau_{21})\mu_{ba}(\tau_1)P(a),
\end{eqnarray}
where
\begin{equation}\label{eq:time.evolution}
I_{ab}(\tau_{21})=\exp\left[-\int_{\tau_1}^{\tau_2}i\epsilon_{a}(\tau)d \tau\right]\exp\left[\int_{\tau_1}^{\tau_2}i\epsilon_{b}(\tau)d \tau\right].
\end{equation}
$\epsilon_a(\tau)$ and $\epsilon_b(\tau)$ are the eigenvalues of the Hamiltonian at time $\tau$. $\mu_{ab}(\tau)$ are the transition dipoles in the adiabatic basis. $P(a)$ is the equilibrium population of state $a$.
When the energies change rapidly, the adiabatic approximation breaks down and nonadiabatic transitions need to be taken into account. 

 In order to simulate the time evolution of the system one would have to diagonalize the vibrational Hamiltonian at short intervals. Such a treatment is therefore only feasible when the number of states is sufficiently small to allow repeated diagonalizations. This may be the case if a subset of vibrational coordinates are of interest and the rest are treated as bath coordinates. It should be noted that treating the vibrational states of the system explicitly only makes sense for high frequency modes, where the excited vibrational states have a low thermal population and only few states need to be considered.

For all methods based on an expansion of the response function in the
coordinates, evaluating the contributing terms gets increasingly more expensive
with the order of the expansion. While the first and the second order derivatives of
the polarizability needed to evaluate the lowest order terms can be handled
rather easily, higher order derivatives get increasingly more difficult to evaluate.
Unless the coordinates can be chosen such that the
expansion may be truncated at some low order, it will be harder to simulate the classical response functions using the expansion, compared with real-space simulation methods.

\section{Conclusions}
We have developed a systematic perturbative method for computing
response functions using an expansion in the nonlinearities,
the effective dipoles, and the anharmonicities. Closed form
expressions for the lower order terms have been derived for the
first-, second- and third order response functions and
applications were made to the 1D and 2D response corresponding to
 third- and fifth order Raman techniques, respectively.

The mode-coupling factorizations presented in this paper provide a unified framework for deriving all
the approximate methods used so far in the simulations of the fifth
order Raman. Our expressions reduce to the instantaneous normal mode as well as the mode-coupling expressions by making additional approximations.
This unified description allows a direct comparison of the various methods and can be used to develop semiclassical expansions. All existing simulations were
compared and connected to the mode-coupling factorization
presented in Sec. II. 

The present green function formalism allows the explicit
incorporation of anhamonicities and gives a rigorous algorithm for
truncating the expansion of the response functions, depending on
the number of derivatives involved. At the same time, quantum
corrections to the response functions may be computed as well.
Mode-coupling theories in $\mathbf{k}$ space~\cite{Denny.2001.PRE.63.065101,Denny.2002.JCP.116.1987,Denny.2002.JCP.116.1979,Cao.2002.JCP.116.3739,Cao.2002.JCP.116.3760}
utilize a non-local polarizability in order to describe response
functions that are local in nature. This leads to the neglect of the
leading order term in both the third and fifth order Raman
response functions; contributions of all terms
involving the first order derivative of the polarizability are missed.

Collective coordinates are likely to have Gaussian statistics. 
Identifying a set of collective coordinates that should allow a relatively simple interpretation of the response is the the key open challenge in the simulation of nonlinear response in the condensed phase.

\section*{Acknowledgement}
The support of the National Institutes of Health grant no. (RO1
GM59230-01A2) and the National Science Foundation grant no.
(CHE-0132571) is gratefully acknowledged. We wish to thank dr. Vladimir Chernyak for most useful discussions. 

\appendix
\section{Superoperator Algebra}
The following relations for the superoperators $Q_-$ and $Q_+$ 
that follow directly from the definitions of the superoperators
(Eq. (\ref{eq:superoperators})) can be used in expand the nonlinear
response in terms of these superoperators.

\begin{eqnarray}\label{eq:quadratic}
(Q^jQ^i)_+ & = & (Q_+^jQ_+^i+\frac{1}{4}Q_-^jQ_-^i) \\
(Q^jQ^i)_- & = & (Q_+^jQ_-^i+Q_-^jQ_+^i)
\end{eqnarray}
\begin{eqnarray}\label{eq:cubed}
(Q^kQ^jQ^i)_+ & = & \frac{1}{4}(4 Q_+^kQ_+^jQ_+^i + Q_+^kQ_-^jQ_-^i + Q_-^kQ_+^jQ_-^i + Q_-^kQ_-^jQ_+^i) \\
(Q^kQ^jQ^i)_- & = & (Q_+^kQ_+^jQ_-^i + Q_+^kQ_-^jQ_+^i + Q_-^kQ_+^jQ_+^i + \frac{1}{4}Q_-^kQ_-^jQ_-^i)
\end{eqnarray}
and so forth.
Using these rules we can express the superoperator
corresponding to an arbitrary product of ordinary operators as a product of
superoperators
\begin{eqnarray}
(Q^{i_n}\cdots Q^{i_1})_-&=& \sum_{\nu_1 \cdots \nu_n}f^-_{\nu_1 \cdots \nu_n}
Q^{i_n}_{\nu_n}\cdots Q^{i_1}_{\nu_1}, \nonumber\\
(Q^{i_n}\cdots Q^{i_1})_+&=& \sum_{\nu_1 \cdots \nu_n}f^+_{\nu_1
\cdots \nu_n} Q^{i_n}_{\nu_n}\cdots Q^{i_1}_{\nu_1},
\end{eqnarray}
where the coefficients $f^+$ and $f^-$ are determined by
application of the rules (Eqs.~\ref{eq:quadratic}-\ref{eq:cubed})
and the $\nu$'s denote either $+$ or $-$.

For a given set of expansion coefficients the n dimensional
response can be expressed in terms of derivatives of the
generating functional.~\cite{Mukamel.submitted}
\begin{eqnarray}\label{eq:monster}
&&R_{m:n_1\cdots n_{n+1}}^{(nD)}(\tau_{n+1},\cdots,\tau_1,\tau'_{m},\cdots,\tau'_{1})
=\nonumber\\ &&
\frac{1+\delta_{m0}}{(m+2)!n_1!\cdots n_{n+1}!}\left(\frac{i}{\hbar}\right)^{m+n}(-1)^m\sum_{j^1\cdots j^{n+1}k^1\cdots k^{m}}
\int_{-\infty}^{\tau_{n+1}}d\tau'_1\cdots\int_{-\infty}^{t_{n+1}}d\tau'_m
\nonumber\\ &&\times
\mu_{j_1^1\cdots j^1_{n_1^1}}^{(n_1)}\cdots\mu_{j_1^{n+1}\cdots j^{n+1}_{n_{n+1}}}^{(n_{n+1})}
V_{k_1^1\cdots k^1_{m_1}}^{m_1}\cdots V_{k_1^m\cdots k_{m_m}^m}\nonumber\\
&&\times\sum_{\nu^1\cdots \nu^{n}} f^-_{\nu^1_1\cdots\nu^1_{n_1}}f^-_{\nu^{n}_1\cdots\nu^{n}_{n_{n}}}\left(
\frac{\partial}{\partial J_{j^{n}_1+}(\tau_{n})}\cdots\frac{\partial}{\partial J_{j^{n}_{n_{n}}+}(\tau_{n+1})}\right)
\nonumber\\ &&\times
\left(\frac{\partial}{\partial J_{j^1_1\nu^1_1}(\tau_1)}\cdots\frac{\partial}{\partial J_{j^1_{n_1}\nu^1_{n_1}}(\tau_1)}\right)\cdots\left(
\frac{\partial}{\partial J_{j^{n}_1\nu^{n}_1}(\tau_{n})}\cdots\frac{\partial}{\partial J_{j^{n}_{n_{n}}\nu^{n}_{n_{n}}}(\tau_{n})}\right)
\nonumber\\ &&\times\sum_{\xi^1\cdots\xi^m}
\left(\frac{\partial}{\partial J_{k^1_1\xi^1_1}(\tau'_1)}\cdots\frac{\partial}{\partial J_{k^1_{m_1}\xi^1_{m_1}}(\tau'_1)}\right)\cdots
\left(\frac{\partial}{\partial J_{k^m_1\xi^m_1}(\tau'_m)}\cdots\frac{\partial}{\partial J_{k^m_{m_m}\xi^m_{m_m}}(\tau'_m)}\right)\nonumber\\
&&\times f^-_{\xi^1_1\cdots\xi^1_{m_1}}\cdots f^-_{\xi^m_1\cdots\xi^m_{m_m}}S \{J(t)\} \bigg|_{J=0}
\end{eqnarray}

\section{$2D$ and $3D$ response to sixth order in the primary coordinates}

The $2D$ and $3D$ response functions are derived in the same way as the $1D$ response function that was given in Eq. (\ref{eq:firstorder}).
The $2D$ response function expanded to sixth order in the primary
coordinates is given by
\begin{equation}\label{eq:fifthorder}
R^{(2D)}=R^{(2D)}_{211}+R^{(2D)}_{222}+R^{(2D)}_{321}+R^{(2D)}_{411}+R^{(2D)}_{3:111}+\cdots,
\end{equation}
where the first term is fourth order in the coordinates
\begin{eqnarray}\label{eq:fourthorder}
R^{(2D)}_{211}&=&\sum_{ijkl}\mu^{(2)}_{ij}\mu^{(1)}_{k}\mu^{(1)}_{l} G^{+-}_{ik}(\tau_{32}) \left( G^{+-}_{jl}(\tau_{31})+ G^{+-}_{jl}(\tau_{21}) \right).
\end{eqnarray}
The remaining contributions are sixth order in the coordinates
\begin{eqnarray}\label{eq:firstsixthorder}
R^{(2D)}_{222}& =& \sum_{ijklmn}\mu^{(2)}_{ij}\mu^{(2)}_{kl}\mu^{(2)}_{mn} G^{+-}_{ik}(\tau_{32}) \left( G^{+-}_{jn}(\tau_{31}) G_{lm}^{++}(\tau_{21}) + G^{+-}_{ln}(\tau_{21}) G_{jm}^{++}(\tau_{31}) \right)
\end{eqnarray}
\begin{eqnarray}\label{eq:secondsixthorder}
R^{(2D)}_{321}& = & \frac{1}{2}
 \sum_{ijklmn}\mu^{(3)}_{ijk}\mu^{(2)}_{lm}\mu^{(1)}_{n} G_{ij}^{++}(0) \left( G^{+-}_{ln}(\tau_{21}) G^{+-}_{km}(\tau_{32}) + G^{+-}_{mk}(\tau_{21}) G^{+-}_{nl}(\tau_{32}) \right)\nonumber\\
& + & \frac{1}{2}
 \sum_{ijklmn}\mu^{(3)}_{ijk}\mu^{(2)}_{lm}\mu^{(1)}_{n} G_{ij}^{++}(0) \left( G^{+-}_{ln}(\tau_{31}) G^{+-}_{mk}(\tau_{32}) + G^{+-}_{mk}(\tau_{31}) G^{+-}_{ln}(\tau_{32}) \right)\nonumber\\
& + &
 \sum_{ijklmn}\mu^{(3)}_{ijk}\mu^{(2)}_{lm}\mu^{(1)}_{n} \left( G_{il}^{++}(\tau_{32}) G^{+-}_{kn}(\tau_{31}) G^{+-}_{jm}(\tau_{32}) +G_{il}^{++}(\tau_{31}) G^{+-}_{km}(\tau_{31}) G^{+-}_{jn}(\tau_{32}) \right)\nonumber\\
& + &
\sum_{ijklmn}\mu^{(3)}_{ijk}\mu^{(2)}_{lm}\mu^{(1)}_{n} \left( G_{li}^{++}(\tau_{32}) G^{+-}_{kn}(\tau_{21}) G^{+-}_{mj}(\tau_{32}) +G_{il}^{++}(\tau_{21}) G^{+-}_{nk}(\tau_{32}) G^{+-}_{jm}(\tau_{21}) \right)\nonumber\\
\end{eqnarray}
\begin{eqnarray}\label{eq:411}
R^{(2D)}_{411}& = & \frac{1}{2} \sum_{ijklmn}\mu^{(4)}_{ijkl}\mu^{(1)}_{m}\mu^{(1)}_{n} G_{ij}^{++}(0) \left( G^{+-}_{ln}(\tau_{31}) G^{+-}_{km}(\tau_{32}) + G^{+-}_{ln}(\tau_{21}) G^{+-}_{mk}(\tau_{32}) \right)
\end{eqnarray}
\begin{equation}\label{eq:3111}
R^{(2D)}_{3:111}=-\sum_{ijklmn}\mu^{(1)}_{i}\mu^{(1)}_{j}\mu^{(1)}_{k} V^{(3)}_{lmn}\int_{-\infty}^{\tau_3} d \tau_{1'} G^{+-}_{il}(\tau_{31'}) G^{+-}_{jm}(\tau_{1'1}) G^{+-}_{kn}(\tau_{1'2})
\end{equation}
The two lowest-order terms contain numerous contributions. As
noted earlier~\cite{Tanimura.1993.JCP.99.9496} all terms in the 2D
response contain nonlinearities either in form of anharmonicities
or higher order derivatives of the dipole operator.

The $3D$ response function expanded to sixth order in the
coordinates is
\begin{eqnarray}
R^{(3D)}& = & \sum_{ijklmn}\mu^{(2)}_{ij}\mu^{(2)}_{kl}\mu^{(1)}_{m}\mu^{(1)}_{n} \Big[G_{ik}^{+-}
(\tau_{43}) \left( G^{+-}_{jm}(\tau_{42}) G^{+-}_{ln}(\tau_{31}) + G^{+-}_{jn}(\tau_{41}) G^{+-}_{lm}(\tau_{32}) \right)\nonumber\\
& + & G_{ik}^{+-}(\tau_{42}) G^{+-}_{jm}(\tau_{43}) G^{+-}_{ln}(\tau_{21})
 +  G_{ik}^{+-}(\tau_{32})G^{+-}_{jn}(\tau_{21}) G^{+-}_{ml}(\tau_{43}) \Big]
\nonumber\\
& + &
\sum_{ijklmn}\mu^{(3)}_{ijk}\mu^{(1)}_{l}\mu^{(1)}_{m}\mu^{(1)}_{n}
G_{im}^{+-}(\tau_{43})G^{+-}_{jn}(\tau_{42})
G^{+-}_{kl}(\tau_{41}).
\end{eqnarray}
Numerous eight order terms exists including an anharmonic term.
They can be readily obtained using the rules outlined earlier and
will not be given here.

\section{The Matrix of Spectral Densities}
The matrix of spectral densities for a system described by the Hamiltonian defined in Eqs. (\ref{2.1t}), (\ref{2.7t}) and (\ref{2.6t}) can be determined by solving the generalized Langevin equation (Eq. (\ref{eq:GLE}).
\cite{Mukamel.1995.B01,PA.121-587,Chernyak.1998.JCP.108.5812}
The generalized Langevin equation can also be written on the form
\begin{equation}\label{eq:GLE2}
M_j\ddot{Q}_j(t)+M_j\Omega_jQ_j(t)+\sum_i\int^{t}_{-\infty}d\tau
\frac{\langle \delta f_j(t-\tau)\delta f_i(0)\rangle}{k_BT}\dot{Q}_i(\tau)=f_j(t)+F_j(t)
\end{equation}
With the external driving force $F_j(t)$ and the rapidly fluctuating force of the bath on the primary coordinate $j$ $f_j(t)$. This force is determined from the system-bath Hamiltonian (Eq.(\ref{2.7t})).
\begin{equation}
f_j=\frac{d H_B}{d Q_j}=-\sum_{\alpha}
(c_{j\alpha}q_{\alpha})
+\sum_{i,\alpha}2\frac{c_{j\alpha}c_{j\alpha}}{2m_{\alpha}\omega_{\alpha}^2}Q_i,
\end{equation}
where the second term is independent of the bath coordinates.

Comparing Eqs. (\ref{eq:GLE}) and (\ref{eq:GLE2}) allows us to identify $\gamma$ and $\Sigma$ as the real and imaginary part of the correlations function of the fluctuating forces $f_j(t)$.
\begin{equation}
\gamma_{ji}(t-\tau)+i\Sigma_{ji}(t-\tau)\equiv\frac{\langle \delta f_j(t-\tau)\delta f_i(0)\rangle}{M_jk_BT}
\end{equation}

Averaging over the bath coordinates and taking the
Fourier transform of the generalized Langevin equation (Eq.(\ref{eq:GLE2})) gives:
\begin{equation}
-M_j\langle\tilde{Q_j}(\omega)\rangle\omega^2+M_j\Omega_j^2\langle\tilde{Q_j}(\omega)\rangle+M_j\sum_i
\left(-i\tilde{\gamma}_{ji}(\omega)+i^2\tilde{\Sigma}_{ji}(\omega)
\right)\omega\langle\tilde{Q_i}(\omega)\rangle=\tilde{F}_j(\omega)
\end{equation}

In matrix form this gives:
\begin{equation}
M(\Omega^2-\omega^2I+\Sigma\omega-i\gamma\omega)\langle\tilde{Q}(\omega)\rangle=\tilde{F}(\omega)
\end{equation}
The matrices ${M}$, ${\Omega}$ and ${I}$ are all diagonal with matrix elements $M_{ij}=\delta_{ij}M_j$, $\Omega_{ij}=\delta_{ij}\Omega_j$ and $I_{ij}=\delta_{ij}$.  $Q$ and $F$ are vectors.

The change in the coordinates induced by an external driving force is
\begin{equation}
\langle\tilde{Q}(\omega)\rangle=\alpha(\omega)\tilde{F}(\omega)=\frac{1}{M(\Omega^2-\omega^2I+\Sigma\omega-i\gamma\omega)}\tilde{F}(\omega),
\end{equation}
which define the susceptibility $\alpha(\omega)$.

The odd part of the spectral density matrix is the imaginary part of the susceptibility:
\begin{eqnarray}\label{ghix}
{C''}(\omega) & = & \textrm{Im}\left(\frac{1}{{M}
({\Omega}^2 +\omega{\Sigma} (\omega)-{I}{\omega}^2 +  i\omega\mathbf{\gamma}(\omega))}\right),
\end{eqnarray}
where the imaginary part of a matrix is:
$
\textrm{Im}\ A={(A-A^{\dagger})}/{2i}
$.

The even part of the spectral density is related to the odd part by the fluctuation-dissipation theorem.
$\gamma$ and $\Sigma$ are determined by the correlation function of the fluctuating forces. The fluctuating force was determined by the derivative of the Hamiltonian.
\begin{equation}
\langle \delta f_j(t)\delta f_i(0)\rangle=\sum_{\alpha\beta}c_{j\alpha}c_{i\beta}
\langle q_{\alpha}(t)q_{\beta}(0)\rangle
\end{equation}
The part of the fluctuating force that does not depend on the bath coordinates vanishes, when averaged over the bath coordinates. 

$\gamma$ is determined by the real part of the time correlation function of the bath coordinates:
\begin{eqnarray}
\gamma_{ij}(t)&=&\sum_{\alpha\beta}\frac{c_{i\alpha}c_{j\beta}}{M_jk_BT}
\textrm{Re}\langle q_{\alpha}(t)q_{\beta}(0)\rangle\\
&=&\sum_{\alpha\beta}\frac{1}{2M_j}\frac{c_{i\alpha}c_{j\beta}}{m_{\alpha}\omega_{\alpha}^2}\delta_{\alpha\beta}\cos(\omega_{\alpha}t)
\end{eqnarray}
$\gamma_{ij}(\omega)$ is the Fourier transform of $\gamma_{ij}(t)$ and is given in Eq. (\ref{ghi1}). 

\newpage

\begin{thebibliography}{75}
\expandafter\ifx\csname natexlab\endcsname\relax\def\natexlab#1{#1}\fi
\expandafter\ifx\csname bibnamefont\endcsname\relax
  \def\bibnamefont#1{#1}\fi
\expandafter\ifx\csname bibfnamefont\endcsname\relax
  \def\bibfnamefont#1{#1}\fi
\expandafter\ifx\csname citenamefont\endcsname\relax
  \def\citenamefont#1{#1}\fi
\expandafter\ifx\csname url\endcsname\relax
  \def\url#1{\texttt{#1}}\fi
\expandafter\ifx\csname urlprefix\endcsname\relax\def\urlprefix{URL }\fi
\providecommand{\bibinfo}[2]{#2}
\providecommand{\eprint}[2][]{\url{#2}}

\bibitem[{\citenamefont{Tanimura and
  Mukamel}(1993)}]{Tanimura.1993.JCP.99.9496}
\bibinfo{author}{\bibfnamefont{Y.}~\bibnamefont{Tanimura}} \bibnamefont{and}
  \bibinfo{author}{\bibfnamefont{S.}~\bibnamefont{Mukamel}},
  \bibinfo{journal}{J. Chem. Phys.} \textbf{\bibinfo{volume}{99}},
  \bibinfo{pages}{9496} (\bibinfo{year}{1993}).

\bibitem[{\citenamefont{Steffen and Duppen}(1996)}]{Steffen.1996.PRL.76.1224}
\bibinfo{author}{\bibfnamefont{T.}~\bibnamefont{Steffen}} \bibnamefont{and}
  \bibinfo{author}{\bibfnamefont{K.}~\bibnamefont{Duppen}},
  \bibinfo{journal}{Phys. Rev. Lett.} \textbf{\bibinfo{volume}{76}},
  \bibinfo{pages}{1224} (\bibinfo{year}{1996}).

\bibitem[{\citenamefont{Steffen and Duppen}(1997)}]{Steffen.1997.JCP.106.3854}
\bibinfo{author}{\bibfnamefont{T.}~\bibnamefont{Steffen}} \bibnamefont{and}
  \bibinfo{author}{\bibfnamefont{K.}~\bibnamefont{Duppen}},
  \bibinfo{journal}{J. Chem. Phys.} \textbf{\bibinfo{volume}{106}},
  \bibinfo{pages}{3854} (\bibinfo{year}{1997}).

\bibitem[{\citenamefont{Steffen and Duppen}(1998)}]{Steffen.1998.CPL.290.229}
\bibinfo{author}{\bibfnamefont{T.}~\bibnamefont{Steffen}} \bibnamefont{and}
  \bibinfo{author}{\bibfnamefont{K.}~\bibnamefont{Duppen}},
  \bibinfo{journal}{Chem. Phys. Lett.} \textbf{\bibinfo{volume}{290}},
  \bibinfo{pages}{229} (\bibinfo{year}{1998}).

\bibitem[{\citenamefont{Tokmakoff et~al.}(1998)\citenamefont{Tokmakoff, Lang,
  Jordanides, and Fleming}}]{Tokmakoff.1998.CP.233.231}
\bibinfo{author}{\bibfnamefont{A.}~\bibnamefont{Tokmakoff}},
  \bibinfo{author}{\bibfnamefont{M.~J.} \bibnamefont{Lang}},
  \bibinfo{author}{\bibfnamefont{X.~L.} \bibnamefont{Jordanides}},
  \bibnamefont{and} \bibinfo{author}{\bibfnamefont{G.~R.}
  \bibnamefont{Fleming}}, \bibinfo{journal}{Chem. Phys.}
  \textbf{\bibinfo{volume}{233}}, \bibinfo{pages}{231} (\bibinfo{year}{1998}).

\bibitem[{\citenamefont{Tokmakoff and
  Fleming}(1997)}]{Tokmakoff.1997.JCP.106.2569}
\bibinfo{author}{\bibfnamefont{A.}~\bibnamefont{Tokmakoff}} \bibnamefont{and}
  \bibinfo{author}{\bibfnamefont{G.~R.} \bibnamefont{Fleming}},
  \bibinfo{journal}{J. Chem. Phys.} \textbf{\bibinfo{volume}{106}},
  \bibinfo{pages}{2569} (\bibinfo{year}{1997}).

\bibitem[{\citenamefont{Tominaga and
  Yoshihara}(1996)}]{Tominaga.1996.JCP.104.4419}
\bibinfo{author}{\bibfnamefont{K.}~\bibnamefont{Tominaga}} \bibnamefont{and}
  \bibinfo{author}{\bibfnamefont{K.}~\bibnamefont{Yoshihara}},
  \bibinfo{journal}{J. Chem. Phys.} \textbf{\bibinfo{volume}{104}},
  \bibinfo{pages}{4419} (\bibinfo{year}{1996}).

\bibitem[{\citenamefont{Blank et~al.}(1999)\citenamefont{Blank, Kaufman, and
  Fleming}}]{Blank.1999.JCP.111.3105}
\bibinfo{author}{\bibfnamefont{D.~A.} \bibnamefont{Blank}},
  \bibinfo{author}{\bibfnamefont{L.~J.} \bibnamefont{Kaufman}},
  \bibnamefont{and} \bibinfo{author}{\bibfnamefont{G.~R.}
  \bibnamefont{Fleming}}, \bibinfo{journal}{J. Chem. Phys.}
  \textbf{\bibinfo{volume}{111}}, \bibinfo{pages}{3105} (\bibinfo{year}{1999}).

\bibitem[{\citenamefont{Golonzka et~al.}(2000)\citenamefont{Golonzka,
  Demirdšven, Khalil, and Tokmakoff}}]{Golonzka.2000.JCP.113.9893}
\bibinfo{author}{\bibfnamefont{O.}~\bibnamefont{Golonzka}},
  \bibinfo{author}{\bibfnamefont{N.}~\bibnamefont{Demirdšven}},
  \bibinfo{author}{\bibfnamefont{M.}~\bibnamefont{Khalil}}, \bibnamefont{and}
  \bibinfo{author}{\bibfnamefont{A.}~\bibnamefont{Tokmakoff}},
  \bibinfo{journal}{J. Chem. Phys.} \textbf{\bibinfo{volume}{113}},
  \bibinfo{pages}{9893} (\bibinfo{year}{2000}).

\bibitem[{\citenamefont{Astinov
  et~al.}(2000{\natexlab{a}})\citenamefont{Astinov, Kubarych, Milne, and
  Miller}}]{Astinov.2000.OL.25.853}
\bibinfo{author}{\bibfnamefont{V.}~\bibnamefont{Astinov}},
  \bibinfo{author}{\bibfnamefont{K.~J.} \bibnamefont{Kubarych}},
  \bibinfo{author}{\bibfnamefont{C.~J.} \bibnamefont{Milne}}, \bibnamefont{and}
  \bibinfo{author}{\bibfnamefont{R.~J.~D.} \bibnamefont{Miller}},
  \bibinfo{journal}{Optics Letters} \textbf{\bibinfo{volume}{25}},
  \bibinfo{pages}{853} (\bibinfo{year}{2000}{\natexlab{a}}).

\bibitem[{\citenamefont{Astinov
  et~al.}(2000{\natexlab{b}})\citenamefont{Astinov, Kubarych, Milne, and
  Dwayne~Miller}}]{Astinov.2000.CPL.327.334}
\bibinfo{author}{\bibfnamefont{A.}~\bibnamefont{Astinov}},
  \bibinfo{author}{\bibfnamefont{K.~J.} \bibnamefont{Kubarych}},
  \bibinfo{author}{\bibfnamefont{C.~J.} \bibnamefont{Milne}}, \bibnamefont{and}
  \bibinfo{author}{\bibfnamefont{R.~J.} \bibnamefont{Dwayne~Miller}},
  \bibinfo{journal}{Chem. Phys. Lett.} \textbf{\bibinfo{volume}{327}},
  \bibinfo{pages}{334} (\bibinfo{year}{2000}{\natexlab{b}}).

\bibitem[{\citenamefont{Kubarych
  et~al.}(2002{\natexlab{a}})\citenamefont{Kubarych, Milne, Lin, Astinov, and
  Miller}}]{Kubarych.2002.JCP.116.2016}
\bibinfo{author}{\bibfnamefont{K.~J.} \bibnamefont{Kubarych}},
  \bibinfo{author}{\bibfnamefont{C.~J.} \bibnamefont{Milne}},
  \bibinfo{author}{\bibfnamefont{S.}~\bibnamefont{Lin}},
  \bibinfo{author}{\bibfnamefont{V.}~\bibnamefont{Astinov}}, \bibnamefont{and}
  \bibinfo{author}{\bibfnamefont{R.~J.~D.} \bibnamefont{Miller}},
  \bibinfo{journal}{J. Chem. Phys.} \textbf{\bibinfo{volume}{116}},
  \bibinfo{pages}{2016} (\bibinfo{year}{2002}{\natexlab{a}}).

\bibitem[{\citenamefont{Kubarych
  et~al.}(2002{\natexlab{b}})\citenamefont{Kubarych, Milne, Lin, and
  Miller}}]{Kubarych.2002.Ultrafast}
\bibinfo{author}{\bibfnamefont{K.~J.} \bibnamefont{Kubarych}},
  \bibinfo{author}{\bibfnamefont{C.~J.} \bibnamefont{Milne}},
  \bibinfo{author}{\bibfnamefont{S.}~\bibnamefont{Lin}}, \bibnamefont{and}
  \bibinfo{author}{\bibfnamefont{R.~J.~D.} \bibnamefont{Miller}}, in
  \emph{\bibinfo{booktitle}{13th International Conference on Ultrafast
  Phenomena}} (\bibinfo{publisher}{Springer Verlag},
  \bibinfo{address}{Vancouver}, \bibinfo{year}{2002}{\natexlab{b}}), vol.
  \bibinfo{volume}{"XIII"} of \emph{\bibinfo{series}{Ultrafast Phenomena}}.

\bibitem[{\citenamefont{Kubarych et~al.}(2003)\citenamefont{Kubarych, Milne,
  and Miller}}]{Kubarych.2003.CPL.369.635}
\bibinfo{author}{\bibfnamefont{K.}~\bibnamefont{Kubarych}},
  \bibinfo{author}{\bibfnamefont{C.~J.} \bibnamefont{Milne}}, \bibnamefont{and}
  \bibinfo{author}{\bibfnamefont{R.~J.~D.} \bibnamefont{Miller}},
  \bibinfo{journal}{Chem. Phys. Lett.} \textbf{\bibinfo{volume}{369}},
  \bibinfo{pages}{635} (\bibinfo{year}{2003}).

\bibitem[{\citenamefont{Kaufman et~al.}(2001)\citenamefont{Kaufman, Blank, and
  Fleming}}]{Kaufman.2001.JCP.114.2312}
\bibinfo{author}{\bibfnamefont{L.~J.} \bibnamefont{Kaufman}},
  \bibinfo{author}{\bibfnamefont{D.~A.} \bibnamefont{Blank}}, \bibnamefont{and}
  \bibinfo{author}{\bibfnamefont{G.~R.} \bibnamefont{Fleming}},
  \bibinfo{journal}{J. Chem. Phys.} \textbf{\bibinfo{volume}{114}},
  \bibinfo{pages}{2312} (\bibinfo{year}{2001}).

\bibitem[{\citenamefont{Kaufman et~al.}(2002)\citenamefont{Kaufman, Heo,
  Ziegler, and Fleming}}]{Kaufman.2002.PRL.88.207402}
\bibinfo{author}{\bibfnamefont{L.~J.} \bibnamefont{Kaufman}},
  \bibinfo{author}{\bibfnamefont{J.}~\bibnamefont{Heo}},
  \bibinfo{author}{\bibfnamefont{L.~D.} \bibnamefont{Ziegler}},
  \bibnamefont{and} \bibinfo{author}{\bibfnamefont{G.~R.}
  \bibnamefont{Fleming}}, \bibinfo{journal}{Phys. Rev. Lett.}
  \textbf{\bibinfo{volume}{88}}, \bibinfo{pages}{207402}
  (\bibinfo{year}{2002}).

\bibitem[{\citenamefont{Ulness et~al.}(1998)\citenamefont{Ulness, Kirkwood, and
  Albrecht}}]{Ulness.1998.JCP.108.3897}
\bibinfo{author}{\bibfnamefont{D.~J.} \bibnamefont{Ulness}},
  \bibinfo{author}{\bibfnamefont{J.~C.} \bibnamefont{Kirkwood}},
  \bibnamefont{and} \bibinfo{author}{\bibfnamefont{A.~C.}
  \bibnamefont{Albrecht}}, \bibinfo{journal}{J. Chem. Phys.}
  \textbf{\bibinfo{volume}{108}}, \bibinfo{pages}{3897} (\bibinfo{year}{1998}).

\bibitem[{\citenamefont{Kirkwood et~al.}(1998)\citenamefont{Kirkwood, Ulness,
  and Albrecht}}]{Kirkwood.1998.CPL.293.417}
\bibinfo{author}{\bibfnamefont{J.~C.} \bibnamefont{Kirkwood}},
  \bibinfo{author}{\bibfnamefont{D.~J.} \bibnamefont{Ulness}},
  \bibnamefont{and} \bibinfo{author}{\bibfnamefont{A.~C.}
  \bibnamefont{Albrecht}}, \bibinfo{journal}{Chem. Phys. Lett.}
  \textbf{\bibinfo{volume}{293}}, \bibinfo{pages}{417} (\bibinfo{year}{1998}).

\bibitem[{\citenamefont{Mukamel et~al.}(1996)\citenamefont{Mukamel, Khidekel,
  and Chernyak}}]{Mukamel.1996.PRE.53.R1}
\bibinfo{author}{\bibfnamefont{S.}~\bibnamefont{Mukamel}},
  \bibinfo{author}{\bibfnamefont{V.}~\bibnamefont{Khidekel}}, \bibnamefont{and}
  \bibinfo{author}{\bibfnamefont{V.}~\bibnamefont{Chernyak}},
  \bibinfo{journal}{Phys. Rev. E} \textbf{\bibinfo{volume}{53}},
  \bibinfo{pages}{R1} (\bibinfo{year}{1996}).

\bibitem[{\citenamefont{Saito and Ohmine}(1998)}]{Saito.1998.JCP.108.240}
\bibinfo{author}{\bibfnamefont{S.}~\bibnamefont{Saito}} \bibnamefont{and}
  \bibinfo{author}{\bibfnamefont{I.}~\bibnamefont{Ohmine}},
  \bibinfo{journal}{J. Chem. Phys.} \textbf{\bibinfo{volume}{108}},
  \bibinfo{pages}{240} (\bibinfo{year}{1998}).

\bibitem[{\citenamefont{Khidekel et~al.}(1996)\citenamefont{Khidekel, Chernyak,
  and Mukamel}}]{Khidekel.1996.B01}
\bibinfo{author}{\bibfnamefont{V.}~\bibnamefont{Khidekel}},
  \bibinfo{author}{\bibfnamefont{V.}~\bibnamefont{Chernyak}}, \bibnamefont{and}
  \bibinfo{author}{\bibfnamefont{S.}~\bibnamefont{Mukamel}}, in
  \emph{\bibinfo{booktitle}{Femtochemistry}}, edited by
  \bibinfo{editor}{\bibfnamefont{M.}~\bibnamefont{Chergui}}
  (\bibinfo{publisher}{World Scientific}, \bibinfo{address}{Singapore},
  \bibinfo{year}{1996}), p. \bibinfo{pages}{570}.

\bibitem[{\citenamefont{Ma and Stratt}(2000)}]{Ma.2000.PRL.85.1004}
\bibinfo{author}{\bibfnamefont{A.}~\bibnamefont{Ma}} \bibnamefont{and}
  \bibinfo{author}{\bibfnamefont{R.~M.} \bibnamefont{Stratt}},
  \bibinfo{journal}{Phys. Rev. Lett.} \textbf{\bibinfo{volume}{85}},
  \bibinfo{pages}{1004} (\bibinfo{year}{2000}).

\bibitem[{\citenamefont{Ma and
  Stratt}(2002{\natexlab{a}})}]{Ma.2002.JCP.116.4962}
\bibinfo{author}{\bibfnamefont{A.}~\bibnamefont{Ma}} \bibnamefont{and}
  \bibinfo{author}{\bibfnamefont{R.~M.} \bibnamefont{Stratt}},
  \bibinfo{journal}{J. Chem. Phys.} \textbf{\bibinfo{volume}{116}},
  \bibinfo{pages}{4962} (\bibinfo{year}{2002}{\natexlab{a}}).

\bibitem[{\citenamefont{Saito and Ohmine}(2002)}]{Saito.2002.PRL.88.207401}
\bibinfo{author}{\bibfnamefont{S.}~\bibnamefont{Saito}} \bibnamefont{and}
  \bibinfo{author}{\bibfnamefont{I.}~\bibnamefont{Ohmine}},
  \bibinfo{journal}{Phys. Rev. Lett.} \textbf{\bibinfo{volume}{88}},
  \bibinfo{pages}{207401} (\bibinfo{year}{2002}).

\bibitem[{\citenamefont{Saito and Ohmine}(submitted)}]{Saito.submitted}
\bibinfo{author}{\bibfnamefont{S.}~\bibnamefont{Saito}} \bibnamefont{and}
  \bibinfo{author}{\bibfnamefont{I.}~\bibnamefont{Ohmine}},
  \bibinfo{journal}{J. Chem. Phys.}  (\bibinfo{year}{submitted}).

\bibitem[{\citenamefont{Jansen et~al.}(2000)\citenamefont{Jansen, Snijders, and
  Duppen}}]{Jansen.2000.JCP.113.307}
\bibinfo{author}{\bibfnamefont{T.~l.~C.} \bibnamefont{Jansen}},
  \bibinfo{author}{\bibfnamefont{J.~G.} \bibnamefont{Snijders}},
  \bibnamefont{and} \bibinfo{author}{\bibfnamefont{K.}~\bibnamefont{Duppen}},
  \bibinfo{journal}{J. Chem. Phys.} \textbf{\bibinfo{volume}{113}},
  \bibinfo{pages}{307} (\bibinfo{year}{2000}).

\bibitem[{\citenamefont{Jansen et~al.}(2001)\citenamefont{Jansen, Snijders, and
  Duppen}}]{Jansen.2001.JCP.114.10910}
\bibinfo{author}{\bibfnamefont{T.~l.~C.} \bibnamefont{Jansen}},
  \bibinfo{author}{\bibfnamefont{J.~G.} \bibnamefont{Snijders}},
  \bibnamefont{and} \bibinfo{author}{\bibfnamefont{K.}~\bibnamefont{Duppen}},
  \bibinfo{journal}{J. Chem. Phys.} \textbf{\bibinfo{volume}{114}},
  \bibinfo{pages}{10910} (\bibinfo{year}{2001}).

\bibitem[{\citenamefont{Jansen}(2002)}]{Jansen.2002.B01}
\bibinfo{author}{\bibfnamefont{T.~l.~C.} \bibnamefont{Jansen}},
  \bibinfo{type}{Ph.d. thesis}, \bibinfo{school}{Rijksuniversiteit Groningen}
  (\bibinfo{year}{2002}).

\bibitem[{\citenamefont{Jansen et~al.}(2003{\natexlab{a}})\citenamefont{Jansen,
  Duppen, and Snijders}}]{Jansen.2003.PRB.67.134206}
\bibinfo{author}{\bibfnamefont{T.~l.~C.} \bibnamefont{Jansen}},
  \bibinfo{author}{\bibfnamefont{K.}~\bibnamefont{Duppen}}, \bibnamefont{and}
  \bibinfo{author}{\bibfnamefont{J.~G.} \bibnamefont{Snijders}},
  \bibinfo{journal}{Phys. Rev. B} \textbf{\bibinfo{volume}{67}},
  \bibinfo{pages}{134206} (\bibinfo{year}{2003}{\natexlab{a}}).

\bibitem[{\citenamefont{Jansen et~al.}(2003{\natexlab{b}})\citenamefont{Jansen,
  Snijders, and Duppen}}]{Jansen.2003.BKCS.submitted}
\bibinfo{author}{\bibfnamefont{T.~l.~C.} \bibnamefont{Jansen}},
  \bibinfo{author}{\bibfnamefont{J.~G.} \bibnamefont{Snijders}},
  \bibnamefont{and} \bibinfo{author}{\bibfnamefont{K.}~\bibnamefont{Duppen}},
  \bibinfo{journal}{Bull. Korean Chem. Soc.} \textbf{\bibinfo{volume}{24}},
  \bibinfo{pages}{In Press} (\bibinfo{year}{2003}{\natexlab{b}}).

\bibitem[{\citenamefont{Stratt}(1995)}]{Stratt.1995.ACR.28.201}
\bibinfo{author}{\bibfnamefont{R.~M.} \bibnamefont{Stratt}},
  \bibinfo{journal}{Acc. Chem. Res.} \textbf{\bibinfo{volume}{28}},
  \bibinfo{pages}{201} (\bibinfo{year}{1995}).

\bibitem[{\citenamefont{Murry et~al.}(1998{\natexlab{a}})\citenamefont{Murry,
  Fourkas, and Keyes}}]{Murry.1998.JCP.109.2814}
\bibinfo{author}{\bibfnamefont{R.~L.} \bibnamefont{Murry}},
  \bibinfo{author}{\bibfnamefont{J.~T.} \bibnamefont{Fourkas}},
  \bibnamefont{and} \bibinfo{author}{\bibfnamefont{T.}~\bibnamefont{Keyes}},
  \bibinfo{journal}{J. Chem. Phys.} \textbf{\bibinfo{volume}{109}},
  \bibinfo{pages}{2814} (\bibinfo{year}{1998}{\natexlab{a}}).

\bibitem[{\citenamefont{Murry et~al.}(1998{\natexlab{b}})\citenamefont{Murry,
  Fourkas, and Keyes}}]{Murry.1998.JCP.109.7913}
\bibinfo{author}{\bibfnamefont{R.~L.} \bibnamefont{Murry}},
  \bibinfo{author}{\bibfnamefont{J.~T.} \bibnamefont{Fourkas}},
  \bibnamefont{and} \bibinfo{author}{\bibfnamefont{T.}~\bibnamefont{Keyes}},
  \bibinfo{journal}{J. Chem. Phys.} \textbf{\bibinfo{volume}{109}},
  \bibinfo{pages}{7913} (\bibinfo{year}{1998}{\natexlab{b}}).

\bibitem[{\citenamefont{Ma and
  Stratt}(2002{\natexlab{b}})}]{Ma.2002.JCP.116.4972}
\bibinfo{author}{\bibfnamefont{A.}~\bibnamefont{Ma}} \bibnamefont{and}
  \bibinfo{author}{\bibfnamefont{R.~M.} \bibnamefont{Stratt}},
  \bibinfo{journal}{J. Chem. Phys.} \textbf{\bibinfo{volume}{116}},
  \bibinfo{pages}{4972} (\bibinfo{year}{2002}{\natexlab{b}}).

\bibitem[{\citenamefont{Keyes}(1997)}]{Keyes.1997.JCP.106.46}
\bibinfo{author}{\bibfnamefont{T.}~\bibnamefont{Keyes}}, \bibinfo{journal}{J.
  Chem. Phys.} \textbf{\bibinfo{volume}{106}}, \bibinfo{pages}{46}
  (\bibinfo{year}{1997}).

\bibitem[{\citenamefont{Keyes and Fourkas}(2000)}]{Keyes.2000.JCP.112.287}
\bibinfo{author}{\bibfnamefont{T.}~\bibnamefont{Keyes}} \bibnamefont{and}
  \bibinfo{author}{\bibfnamefont{J.~T.} \bibnamefont{Fourkas}},
  \bibinfo{journal}{J. Chem. Phys.} \textbf{\bibinfo{volume}{112}},
  \bibinfo{pages}{287} (\bibinfo{year}{2000}).

\bibitem[{\citenamefont{Ji et~al.}(2000)\citenamefont{Ji, Alhborn, Space,
  Moore, Zhou, Constantine, and Ziegler}}]{Ji.2000.JCP.112.4186}
\bibinfo{author}{\bibfnamefont{X.}~\bibnamefont{Ji}},
  \bibinfo{author}{\bibfnamefont{H.}~\bibnamefont{Alhborn}},
  \bibinfo{author}{\bibfnamefont{B.}~\bibnamefont{Space}},
  \bibinfo{author}{\bibfnamefont{P.~B.} \bibnamefont{Moore}},
  \bibinfo{author}{\bibfnamefont{Y.}~\bibnamefont{Zhou}},
  \bibinfo{author}{\bibfnamefont{S.}~\bibnamefont{Constantine}},
  \bibnamefont{and} \bibinfo{author}{\bibfnamefont{L.~D.}
  \bibnamefont{Ziegler}}, \bibinfo{journal}{J. Chem. Phys.}
  \textbf{\bibinfo{volume}{112}}, \bibinfo{pages}{4186} (\bibinfo{year}{2000}).

\bibitem[{\citenamefont{Leegwater and Mukamel}(1995)}]{JCP.102-2365}
\bibinfo{author}{\bibfnamefont{J.~A.} \bibnamefont{Leegwater}}
  \bibnamefont{and} \bibinfo{author}{\bibfnamefont{S.}~\bibnamefont{Mukamel}},
  \bibinfo{journal}{J. Chem. Phys.} \textbf{\bibinfo{volume}{102}},
  \bibinfo{pages}{2365} (\bibinfo{year}{1995}).

\bibitem[{\citenamefont{Chernyak and
  Mukamel}(1998)}]{Chernyak.1998.JCP.108.5812}
\bibinfo{author}{\bibfnamefont{V.}~\bibnamefont{Chernyak}} \bibnamefont{and}
  \bibinfo{author}{\bibfnamefont{S.}~\bibnamefont{Mukamel}},
  \bibinfo{journal}{J. Chem. Phys.} \textbf{\bibinfo{volume}{108}},
  \bibinfo{pages}{5812} (\bibinfo{year}{1998}).

\bibitem[{\citenamefont{Mukamel}(1995)}]{Mukamel.1995.B01}
\bibinfo{author}{\bibfnamefont{S.}~\bibnamefont{Mukamel}},
  
\emph{\bibinfo{title}{Principles of Nonlinear Optical Spectroscopy}}
  (\bibinfo{publisher}{Oxford University Press}, \bibinfo{address}{New York},
  \bibinfo{year}{1995}).

\bibitem[{\citenamefont{Mukamel et~al.}(1999)\citenamefont{Mukamel,
  Piryatinski, and Chernyak}}]{Mukamel.1999.JCP.110.1711}
\bibinfo{author}{\bibfnamefont{S.}~\bibnamefont{Mukamel}},
  \bibinfo{author}{\bibfnamefont{A.}~\bibnamefont{Piryatinski}},
  \bibnamefont{and} \bibinfo{author}{\bibfnamefont{V.}~\bibnamefont{Chernyak}},
  \bibinfo{journal}{J. Chem. Phys.} \textbf{\bibinfo{volume}{110}},
  \bibinfo{pages}{1711} (\bibinfo{year}{1999}).

\bibitem[{\citenamefont{Palese et~al.}(1994)\citenamefont{Palese, Buontempo,
  Schilling, Lotshaw, Tanimura, Mukamel, and Miller}}]{JPC.98-12466}
\bibinfo{author}{\bibfnamefont{S.}~\bibnamefont{Palese}},
  \bibinfo{author}{\bibfnamefont{J.~T.} \bibnamefont{Buontempo}},
  \bibinfo{author}{\bibfnamefont{L.}~\bibnamefont{Schilling}},
  \bibinfo{author}{\bibfnamefont{W.~T.} \bibnamefont{Lotshaw}},
  \bibinfo{author}{\bibfnamefont{Y.}~\bibnamefont{Tanimura}},
  \bibinfo{author}{\bibfnamefont{S.}~\bibnamefont{Mukamel}}, \bibnamefont{and}
  \bibinfo{author}{\bibfnamefont{R.~J.~D.} \bibnamefont{Miller}},
  \bibinfo{journal}{J Phys Chem-Us} \textbf{\bibinfo{volume}{98}},
  \bibinfo{pages}{12466} (\bibinfo{year}{1994}).

\bibitem[{\citenamefont{Palese et~al.}(1996)\citenamefont{Palese, Mukamel,
  Miller, and Lotshaw}}]{JPC.100-10380}
\bibinfo{author}{\bibfnamefont{S.}~\bibnamefont{Palese}},
  \bibinfo{author}{\bibfnamefont{S.}~\bibnamefont{Mukamel}},
  \bibinfo{author}{\bibfnamefont{R.~J.~D.} \bibnamefont{Miller}},
  \bibnamefont{and} \bibinfo{author}{\bibfnamefont{W.~T.}
  \bibnamefont{Lotshaw}}, \bibinfo{journal}{J Phys Chem-Us}
  \textbf{\bibinfo{volume}{100}}, \bibinfo{pages}{10380}
  (\bibinfo{year}{1996}).

\bibitem[{\citenamefont{Fried and Mukamel}(1993)}]{ACP.84-435}
\bibinfo{author}{\bibfnamefont{L.~E.} \bibnamefont{Fried}} \bibnamefont{and}
  \bibinfo{author}{\bibfnamefont{S.}~\bibnamefont{Mukamel}},
  \bibinfo{journal}{Advanced Chemical Physics} \textbf{\bibinfo{volume}{84}},
  \bibinfo{pages}{435} (\bibinfo{year}{1993}).

\bibitem[{\citenamefont{Schofield et~al.}(1992)\citenamefont{Schofield, Lim,
  and Oppenheim}}]{Schofield.1992.PA.181.89}
\bibinfo{author}{\bibfnamefont{J.}~\bibnamefont{Schofield}},
  \bibinfo{author}{\bibfnamefont{J.}~\bibnamefont{Lim}}, \bibnamefont{and}
  \bibinfo{author}{\bibfnamefont{I.}~\bibnamefont{Oppenheim}},
  \bibinfo{journal}{Physica A} \textbf{\bibinfo{volume}{181}},
  \bibinfo{pages}{89} (\bibinfo{year}{1992}).

\bibitem[{\citenamefont{Bouchaud et~al.}(1996)\citenamefont{Bouchaud,
  Cugliandolo, Kurchan, and Mezard}}]{Bouchaud.1996.PA.226.243}
\bibinfo{author}{\bibfnamefont{J.-P.} \bibnamefont{Bouchaud}},
  \bibinfo{author}{\bibfnamefont{L.}~\bibnamefont{Cugliandolo}},
  \bibinfo{author}{\bibfnamefont{J.}~\bibnamefont{Kurchan}}, \bibnamefont{and}
  \bibinfo{author}{\bibfnamefont{M.}~\bibnamefont{Mezard}},
  \bibinfo{journal}{Physica A} \textbf{\bibinfo{volume}{226}},
  \bibinfo{pages}{243} (\bibinfo{year}{1996}).

\bibitem[{\citenamefont{Ronis}(1979)}]{PA.99-403}
\bibinfo{author}{\bibfnamefont{D.}~\bibnamefont{Ronis}},
  \bibinfo{journal}{Physica A} \textbf{\bibinfo{volume}{99}},
  \bibinfo{pages}{403} (\bibinfo{year}{1979}).

\bibitem[{\citenamefont{Gotze}(1989)}]{Gotze.1989.B01}
\bibinfo{author}{\bibfnamefont{W.}~\bibnamefont{Gotze}}, in
  \emph{\bibinfo{booktitle}{Liquids, Freezing and Glass Transition}}, edited by
  \bibinfo{editor}{\bibfnamefont{J.~P.} \bibnamefont{Hansen}},
  \bibinfo{editor}{\bibfnamefont{D.}~\bibnamefont{Levesque}}, \bibnamefont{and}
  \bibinfo{editor}{\bibfnamefont{D.}~\bibnamefont{Zinn-Justin}}
  (\bibinfo{publisher}{Les Houches}, \bibinfo{address}{North Holland,
  Amsterdam}, \bibinfo{year}{1989}).

\bibitem[{\citenamefont{Gotze and Sjogren}(1992)}]{RPP.55-241}
\bibinfo{author}{\bibfnamefont{W.}~\bibnamefont{Gotze}} \bibnamefont{and}
  \bibinfo{author}{\bibfnamefont{L.}~\bibnamefont{Sjogren}},
  \bibinfo{journal}{Rep. Prog. Phys.} \textbf{\bibinfo{volume}{55}},
  \bibinfo{pages}{241} (\bibinfo{year}{1992}).

\bibitem[{\citenamefont{Ronis}(1981)}]{PA.107-25}
\bibinfo{author}{\bibfnamefont{D.}~\bibnamefont{Ronis}},
  \bibinfo{journal}{Physica A} \textbf{\bibinfo{volume}{107}},
  \bibinfo{pages}{25} (\bibinfo{year}{1981}).

\bibitem[{\citenamefont{Mukamel et~al.}(1983)\citenamefont{Mukamel, Stern, and
  Ronis}}]{PRL.50-590}
\bibinfo{author}{\bibfnamefont{S.}~\bibnamefont{Mukamel}},
  \bibinfo{author}{\bibfnamefont{P.~S.} \bibnamefont{Stern}}, \bibnamefont{and}
  \bibinfo{author}{\bibfnamefont{D.}~\bibnamefont{Ronis}},
  \bibinfo{journal}{Phys. Rev. Lett.} \textbf{\bibinfo{volume}{50}},
  \bibinfo{pages}{590} (\bibinfo{year}{1983}).

\bibitem[{\citenamefont{Denny and Reichman}(2001)}]{Denny.2001.PRE.63.065101}
\bibinfo{author}{\bibfnamefont{R.~A.} \bibnamefont{Denny}} \bibnamefont{and}
  \bibinfo{author}{\bibfnamefont{D.~R.} \bibnamefont{Reichman}},
  \bibinfo{journal}{Phys. Rev. E} \textbf{\bibinfo{volume}{63}},
  \bibinfo{pages}{065101(R)} (\bibinfo{year}{2001}).

\bibitem[{\citenamefont{Denny and
  Reichman}(2002{\natexlab{a}})}]{Denny.2002.JCP.116.1987}
\bibinfo{author}{\bibfnamefont{R.~A.} \bibnamefont{Denny}} \bibnamefont{and}
  \bibinfo{author}{\bibfnamefont{D.~R.} \bibnamefont{Reichman}},
  \bibinfo{journal}{J. Chem. Phys.} \textbf{\bibinfo{volume}{116}},
  \bibinfo{pages}{1987} (\bibinfo{year}{2002}{\natexlab{a}}).

\bibitem[{\citenamefont{Denny and
  Reichman}(2002{\natexlab{b}})}]{Denny.2002.JCP.116.1979}
\bibinfo{author}{\bibfnamefont{R.~A.} \bibnamefont{Denny}} \bibnamefont{and}
  \bibinfo{author}{\bibfnamefont{D.~R.} \bibnamefont{Reichman}},
  \bibinfo{journal}{J. Chem. Phys.} \textbf{\bibinfo{volume}{116}},
  \bibinfo{pages}{1979} (\bibinfo{year}{2002}{\natexlab{b}}).

\bibitem[{\citenamefont{Cao et~al.}(2002{\natexlab{a}})\citenamefont{Cao, Wu,
  and Yang}}]{Cao.2002.JCP.116.3739}
\bibinfo{author}{\bibfnamefont{J.}~\bibnamefont{Cao}},
  \bibinfo{author}{\bibfnamefont{J.}~\bibnamefont{Wu}}, \bibnamefont{and}
  \bibinfo{author}{\bibfnamefont{S.}~\bibnamefont{Yang}}, \bibinfo{journal}{J.
  Chem. Phys.} \textbf{\bibinfo{volume}{116}}, \bibinfo{pages}{3739}
  (\bibinfo{year}{2002}{\natexlab{a}}).

\bibitem[{\citenamefont{Cao et~al.}(2002{\natexlab{b}})\citenamefont{Cao, Wu,
  and Yang}}]{Cao.2002.JCP.116.3760}
\bibinfo{author}{\bibfnamefont{J.}~\bibnamefont{Cao}},
  \bibinfo{author}{\bibfnamefont{J.}~\bibnamefont{Wu}}, \bibnamefont{and}
  \bibinfo{author}{\bibfnamefont{S.}~\bibnamefont{Yang}}, \bibinfo{journal}{J.
  Chem. Phys.} \textbf{\bibinfo{volume}{116}}, \bibinfo{pages}{3760}
  (\bibinfo{year}{2002}{\natexlab{b}}).

\bibitem[{\citenamefont{Kim and Keyes}(2002)}]{Kim.2002.PRE.65.061102}
\bibinfo{author}{\bibfnamefont{J.}~\bibnamefont{Kim}} \bibnamefont{and}
  \bibinfo{author}{\bibfnamefont{T.}~\bibnamefont{Keyes}},
  \bibinfo{journal}{Phys. Rev. E} \textbf{\bibinfo{volume}{65}},
  \bibinfo{pages}{061102} (\bibinfo{year}{2002}).

\bibitem[{\citenamefont{Zwanzig}(2001)}]{Zwanzig.2001.B01}
\bibinfo{author}{\bibfnamefont{R.}~\bibnamefont{Zwanzig}},
  \emph{\bibinfo{title}{Nonequilibrium Statistical Mechanics}}
  (\bibinfo{publisher}{Oxford University Press}, \bibinfo{address}{Oxford},
  \bibinfo{year}{2001}).

\bibitem[{\citenamefont{Zon and Schofield}(2002)}]{Zon.2002.PRE.65.011106}
\bibinfo{author}{\bibfnamefont{R.~v.} \bibnamefont{Zon}} \bibnamefont{and}
  \bibinfo{author}{\bibfnamefont{J.}~\bibnamefont{Schofield}},
  \bibinfo{journal}{Phys. Rev. E} \textbf{\bibinfo{volume}{65}},
  \bibinfo{pages}{011106} (\bibinfo{year}{2002}).

\bibitem[{\citenamefont{Mukamel}(in press)}]{Mukamel.submitted}
\bibinfo{author}{\bibfnamefont{S.}~\bibnamefont{Mukamel}},
  \bibinfo{journal}{Phys. Rev. E}  (\bibinfo{year}{in press}).

\bibitem[{\citenamefont{Chernyak et~al.}(1995)\citenamefont{Chernyak, Wang, and
  Mukamel}}]{PR.263-213}
\bibinfo{author}{\bibfnamefont{V.}~\bibnamefont{Chernyak}},
  \bibinfo{author}{\bibfnamefont{N.~J.} \bibnamefont{Wang}}, \bibnamefont{and}
  \bibinfo{author}{\bibfnamefont{S.}~\bibnamefont{Mukamel}},
  \bibinfo{journal}{Phys. Rep.} \textbf{\bibinfo{volume}{263}},
  \bibinfo{pages}{213} (\bibinfo{year}{1995}).

\bibitem[{\citenamefont{Chernyak and
  Mukamel}(1996)}]{Chernyak.1996.JCP.105.4565}
\bibinfo{author}{\bibfnamefont{V.}~\bibnamefont{Chernyak}} \bibnamefont{and}
  \bibinfo{author}{\bibfnamefont{S.}~\bibnamefont{Mukamel}},
  \bibinfo{journal}{J. Chem. Phys.} \textbf{\bibinfo{volume}{105}},
  \bibinfo{pages}{4565} (\bibinfo{year}{1996}).

\bibitem[{\citenamefont{Caldeira and Leggett}(1983)}]{PA.121-587}
\bibinfo{author}{\bibfnamefont{A.~O.} \bibnamefont{Caldeira}} \bibnamefont{and}
  \bibinfo{author}{\bibfnamefont{A.~J.} \bibnamefont{Leggett}},
  \bibinfo{journal}{Physica A} \textbf{\bibinfo{volume}{121}},
  \bibinfo{pages}{587} (\bibinfo{year}{1983}).

\bibitem[{\citenamefont{Dellago and
  Mukamel}(2003)}]{Dellago.2003.PRE.67.035205}
\bibinfo{author}{\bibfnamefont{C.}~\bibnamefont{Dellago}} \bibnamefont{and}
  \bibinfo{author}{\bibfnamefont{S.}~\bibnamefont{Mukamel}},
  \bibinfo{journal}{Phys. Rev. E} \textbf{\bibinfo{volume}{67}},
  \bibinfo{pages}{035205} (\bibinfo{year}{2003}).

\bibitem[{\citenamefont{Brack and Bhaduri}(1997)}]{Brack.1997.B01.C}
\bibinfo{author}{\bibfnamefont{M.}~\bibnamefont{Brack}} \bibnamefont{and}
  \bibinfo{author}{\bibfnamefont{R.~K.} \bibnamefont{Bhaduri}}, in
  \emph{\bibinfo{booktitle}{Semiclassical Physics}}
  (\bibinfo{publisher}{Addison-Wesley Publishing Company, Inc.},
  \bibinfo{address}{Reading, Massachusetts}, \bibinfo{year}{1997}),
  vol.~\bibinfo{volume}{96} of \emph{\bibinfo{series}{Frontiers in Physics}}.

\bibitem[{\citenamefont{Williams and
  Loring}(2000)}]{Williams.2000.JCP.113.10651}
\bibinfo{author}{\bibfnamefont{R.~B.} \bibnamefont{Williams}} \bibnamefont{and}
  \bibinfo{author}{\bibfnamefont{R.~F.} \bibnamefont{Loring}},
  \bibinfo{journal}{J. Chem. Phys.} \textbf{\bibinfo{volume}{113}},
  \bibinfo{pages}{10651} (\bibinfo{year}{2000}).

\bibitem[{\citenamefont{Geiger and Ladanyi}(1987)}]{Geiger.1987.JCP.87.191}
\bibinfo{author}{\bibfnamefont{L.~C.} \bibnamefont{Geiger}} \bibnamefont{and}
  \bibinfo{author}{\bibfnamefont{B.~M.} \bibnamefont{Ladanyi}},
  \bibinfo{journal}{J. Chem. Phys.} \textbf{\bibinfo{volume}{87}},
  \bibinfo{pages}{191} (\bibinfo{year}{1987}).

\bibitem[{\citenamefont{Bender and Orzag}(1978)}]{Bender.1978.B01}
\bibinfo{author}{\bibfnamefont{C.~M.} \bibnamefont{Bender}} \bibnamefont{and}
  \bibinfo{author}{\bibfnamefont{S.~A.} \bibnamefont{Orzag}},
  \emph{\bibinfo{title}{Advanced Mathematical Methods for Scientists and
  Engeneers}} (\bibinfo{publisher}{McGraw-Hill}, \bibinfo{address}{New York},
  \bibinfo{year}{1978}).

\bibitem[{\citenamefont{Hanamura and Mukamel}(1986)}]{JOSAB.3-1124}
\bibinfo{author}{\bibfnamefont{E.}~\bibnamefont{Hanamura}} \bibnamefont{and}
  \bibinfo{author}{\bibfnamefont{S.}~\bibnamefont{Mukamel}},
  \bibinfo{journal}{J. Opt. Soc. Am. B} \textbf{\bibinfo{volume}{3}},
  \bibinfo{pages}{1124} (\bibinfo{year}{1986}).

\bibitem[{\citenamefont{Mukamel and Hanamura}(1986)}]{PRA.33-1099}
\bibinfo{author}{\bibfnamefont{S.}~\bibnamefont{Mukamel}} \bibnamefont{and}
  \bibinfo{author}{\bibfnamefont{E.}~\bibnamefont{Hanamura}},
  \bibinfo{journal}{Phys. Rev. A} \textbf{\bibinfo{volume}{33}},
  \bibinfo{pages}{1099} (\bibinfo{year}{1986}).

\bibitem[{\citenamefont{Mukamel}(1988)}]{ACP.70-165}
\bibinfo{author}{\bibfnamefont{S.}~\bibnamefont{Mukamel}},
  \bibinfo{journal}{Adv. Chem. Phys.} \textbf{\bibinfo{volume}{70}},
  \bibinfo{pages}{165} (\bibinfo{year}{1988}).

\bibitem[{\citenamefont{Weiss}(1993)}]{Weiss.1993.B01}
\bibinfo{author}{\bibfnamefont{U.}~\bibnamefont{Weiss}},
  \emph{\bibinfo{title}{Quantum Dissipative Systems}}
  (\bibinfo{publisher}{World Scientific}, \bibinfo{address}{Singapore},
  \bibinfo{year}{1993}).

\bibitem[{\citenamefont{Griffith}(1995)}]{Griffith.1995.B01}
\bibinfo{author}{\bibfnamefont{D.~J.} \bibnamefont{Griffith}},
  \emph{\bibinfo{title}{Introduction to quantum mechanics}}
  (\bibinfo{publisher}{Prentice Hall}, \bibinfo{address}{Englewood Cliffs,
  N.J.}, \bibinfo{year}{1995}).

\bibitem[{\citenamefont{Shapere and Wilczek}(1989)}]{Shapere.1989.B01}
\bibinfo{editor}{\bibfnamefont{A.}~\bibnamefont{Shapere}} \bibnamefont{and}
  \bibinfo{editor}{\bibfnamefont{F.}~\bibnamefont{Wilczek}}, eds.,
  \emph{\bibinfo{title}{Geometric Phases in Physics}}
  (\bibinfo{publisher}{World Scientific}, \bibinfo{address}{Singapore},
  \bibinfo{year}{1989}).

\bibitem[{\citenamefont{Sakurai}(1978)}]{Sakurai.1978.B01}
\bibinfo{author}{\bibfnamefont{J.~J.} \bibnamefont{Sakurai}},
  \emph{\bibinfo{title}{Advanced quantum mechanics}}
  (\bibinfo{publisher}{Addison-Wesley Pub. Co. Inc.},
  \bibinfo{address}{Reading, Mass.}, \bibinfo{year}{1978}).

\end{thebibliography}

\end{document}